\documentclass[12pt]{article}

\usepackage{appendix}
\usepackage{epsfig}
\usepackage{capt-of}

\usepackage[activeacute,english]{babel}
\usepackage{latexsym}

\newcommand{\bn}{\begin{eqnarray}}
\newcommand{\en}{\end{eqnarray}}
\newcommand{\be}{\begin{equation}}
\newcommand{\ee}{\end{equation}}

\newcommand{\bea}{\begin{eqnarray}}
\newcommand{\eea}{\end{eqnarray}}

\newcommand{\lsim}{\mbox{\raisebox{-.9ex}{~$\stackrel{\mbox{$<$}}{\sim}$~}}}
\newcommand{\gsim}{\mbox{\raisebox{-.9ex}{~$\stackrel{\mbox{$>$}}{\sim}$~}}}

\begin{document}

\title{\begin{flushright}
\normalsize PI/UAN-2008-261FT
\end{flushright}
\vspace{5mm}
{\bf On the Issue of the $\zeta$ Series Convergence and Loop Corrections in the Generation of Observable Primordial Non-Gaussianity in Slow-Roll Inflation. Part I: the Bispectrum}}



\author{
\hspace{-8mm}\textbf{Heiner R. S. Cogollo$^{1,}$\thanks{e-mail: \texttt{heiner.sarmiento@ciencias.uis.edu.co}}, Yeinzon Rodr\'{\i}guez$^{1,2,}$\thanks{e-mail:
\texttt{yeinzon.rodriguez@uan.edu.co}}, C\'esar A. Valenzuela-Toledo$^{1,}$\thanks{e-mail: \texttt{cavalto@ciencias.uis.edu.co} }} \\ \\
\hspace{-12mm}\textit{$^1$Escuela de F\'{\i}sica, Universidad Industrial de Santander,}  \\
\hspace{-12mm}\textit{Ciudad Universitaria, Bucaramanga, Colombia} \\
\hspace{-12mm}\textit{and} \\
\hspace{-12mm}\textit{$^2$Centro de Investigaciones, Universidad Antonio Nari\~no,}\\
\hspace{-12mm}\textit{Cra 3 Este \# 47A-15, Bogot\'a D.C., Colombia}\\
}

\maketitle

\begin{abstract}

\noindent 
We show in this paper that it is possible to attain very high, {\it including observable}, values for the level of non-gaussianity $f_{NL}$ associated with the bispectrum $B_\zeta$ of the primordial curvature perturbation $\zeta$, in a subclass of small-field {\it slow-roll} models of inflation with canonical kinetic terms. Such a result is obtained by taking care of loop corrections both in the spectrum $P_\zeta$ and the bispectrum $B_\zeta$. Sizeable values for $f_{NL}$ arise even if $\zeta$ is generated during inflation. Five issues are considered when constraining the available parameter space:  1. we must ensure that we are in a perturbative regime so that the $\zeta$ series expansion, and its truncation, are valid. 2. we must apply the correct condition for the (possible) loop dominance in $B_\zeta$ and/or $P_\zeta$. 3. we must satisfy the spectrum normalisation condition.  4. we must satisfy the spectral tilt constraint. 5. we must have enough inflation to solve the horizon problem.  
\end{abstract}

\section{Introduction}

Since COBE \cite{cobe} discovered and mapped the anisotropies in the temperature of the cosmic microwave background radiation \cite{smooth}, many balloon and satellite experiments have refined
the measurements of such anisotropies, reaching up to now an amazing combined precision. The COBE sequel has continued with the WMAP satellite \cite{wmap} which has been able to measure the temperature angular power spectrum up to the third peak with unprecedent precision \cite{hinshaw}, and increase the level of sensitivity to primordial non-gaussianity in the bispectrum by two orders of magnitude compared to COBE \cite{komatsu2,komatsu1}.  The next-to-WMAP satellite, PLANCK \cite{planck}, whose launch is programmed for October 2008, is expected to precisely measure the temperature angular power spectrum up to the eighth peak \cite{planck1}, and improve the level of sensitivity to primordial non-gaussianity in the bispectrum by one order of magnitude compared to WMAP \cite{komatsu}.

Because of the progressive improvement in the accuracy of the satellite measurements 
described above, it is pertinent to study cosmological inflationary models that generate significant (and observable) levels of non-gaussianity. An interesting way to address the problem involves the $\delta\textit{N}$ formalism \cite{starobinsky,sasaki2,lyth4}, which can be employed to give the levels of non-gaussianity $f_{NL}$ \cite{lyth2} and $\tau_{NL}$ \cite{boubekeur1,alabidi2} in the bispectrum $B_\zeta$ and trispectrum $T_\zeta$ of the primordial curvature perturbation $\zeta$ respectively. Such non-gaussianity levels are given, for slow-roll inflationary models, in terms of the local evolution of the universe under consideration, as well as of the $n$-point correlators, evaluated a few Hubble times after horizon exit, of the perturbations $\delta\phi_{i}$ in the scalar fields that determine the dynamics of such a universe during inflation.  

In the $\delta\textit{N}$ formalism for slow-roll inflationary models, the primordial curvature perturbation $\zeta(\textbf{x},t)$ is written as a Taylor series in the scalar field perturbations $\delta\phi_{i}(\textbf{x},t_\star)$, evaluated a few Hubble times after horizon exit,
 \begin{eqnarray}
\zeta(t,\textbf{x})&=&\sum_{i}N_{i}(t)\delta\phi_{i}(\textbf{x},t_\star) - \sum_{i}N_{i}(t) \langle\delta\phi_{i}(\textbf{x},t_\star)\rangle + \nonumber \\
&&+\frac{1}{2}\sum_{ij}N_{ij}(t)\delta\phi_{i}(\textbf{x},t_\star)\delta\phi_{j}(\textbf{x},t_\star)-\frac{1}{2}\sum_{ij}N_{ij}(t) \langle\delta\phi_{i}(\textbf{x},t_\star)\delta\phi_{j}(\textbf{x},t_\star)\rangle + \nonumber \\
&&+\frac{1}{3!}\sum_{ijk}N_{ijk}(t)\delta\phi_{i}(\textbf{x},t_\star)\delta\phi_{j}(\textbf{x},t_\star)\delta\phi_{k}(\textbf{x},t_\star) - \frac{1}{3!}\sum_{ijk}N_{ijk}(t)\langle\delta\phi_{i}(\textbf{x},t_\star)\delta\phi_{j}(\textbf{x},t_\star)\delta\phi_{k}(\textbf{x},t_\star)\rangle + \nonumber \\
&&+...\;,
\end{eqnarray}
where the brackets mean spatial averages, $N$ is the amount of inflation (or number of e-folds) from a bit later (in Hubble times) than the time when the cosmologically relevant scales exit the horizon and until the time at which one wishes to calculate $\zeta$, and $N_{i}\equiv\frac{\partial N}{\partial\phi_{i}}$,
$N_{ij}\equiv\frac{\partial^{2}N}{\partial\phi_{i}\partial\phi_{j}}$, and so on. It is in this way that the correlation functions of $\zeta$ (for instance, $\langle\zeta_{\bf k_{1}}\zeta_{\bf k_{2}}\zeta_{\bf k_{3}}\rangle$) can be obtained in terms of series, as often happens in Quantum Field Theory where the probability amplitude is a series whose possible truncation at any desired order is determined by the coupling constants of the theory. A highly relevant question is that of whether the series for $\delta N$ converges in cosmological perturbation theory and whether it is possible in addition to find some quantities that determine the possible truncation of the series, which in this sense would be analogous to the coupling constants in Quantum Field Theory.  In general such  quantities 
will depend on the specific inflationary model; the series then cannot be simply truncated at some order until one is sure that it does indeed converge, and besides, one has to be careful not to forget any term that may be leading in the series even if it is of higher order in the coupling constant. This issue has not been investigated in the present literature, and generally the series has been truncated to second- or third-order neglecting in addition terms that could be the leading ones \cite{sasaki2,lyth2,boubekeur1,alabidi2,zaballa,alabidi1,vernizzi,battefeld,yokoyama1,yokoyama2,seery3,byrnes2}.

The most studied and popular inflationary models nowadays are those of the slow-roll variety with canonical kinetic terms \cite{liddle,lyth5,lyth6}, because of their simplicity and because they easily satisfy the spectral index requirements for the generation of large-scale structures. One of the usual predictions from inflation and the theory of cosmological perturbations is that the levels of non-gaussianity in the primordial perturbations are expected to be unobservably small when considering this class of models \cite{zaballa,vernizzi,battefeld,yokoyama1,seery3,maldacena,seery7,seery5,li,seery4}\footnote{One possible exception is the two-field slow-roll model analyzed in Ref. \cite{alabidi1} (see also Refs. \cite{bernardeu1,bernardeu2}) where {\it observable, of order one, values for} $f_{NL}$ are generated for
a reduced window parameter associated with
the initial field values when taking into account only the tree-level terms in both $P_\zeta$ and $B_\zeta$. However, such a result seems to be incompatible with the general expectation, proved in Ref. \cite{vernizzi}, of $f_{NL}$ being of order the slow-roll parameters, and {\it in consequence unobservable}, for two-field slow-roll models with separable potential when considering only the tree-level terms both in $P_\zeta$ and $B_\zeta$.  The origin of the discrepancy could be understood by conjecturing that the trajectory in field space, for the models in Refs. \cite{alabidi1,bernardeu1,bernardeu2}, seems to be sharply curved, being quite near a saddle point; such a condition is required, according to Ref. \cite{vernizzi}, to generate $f_{NL} \sim \mathcal{O}(1)$. \label{laila}}.
This fact
leads us to analyze the cosmological perturbations in the framework of first-order cosmological perturbation theory.  Non-gaussian characteristics are then suppressed since the non-linearities in the inflaton potential and in the metric perturbations are not taken into account.  The non-gaussian characteristics are actually present and they are made explicit if second-order \cite{lyth3} or higher-order corrections are considered.

The whole literature that encompasses the slow-roll inflationary models with canonical kinetic terms reports that the non-gaussianity level $f_{NL}$ is expected to be very small, being of the order of the slow-roll parameters $\epsilon_i$ and $\eta_i$, ($\epsilon_i, |\eta_i| \ll 1$) \cite{vernizzi,battefeld,yokoyama1,maldacena,seery7}. These works have not taken into account either the convergence of the series for $\zeta$ nor the possibility that loop corrections dominate over the tree level ones in the $n$-point correlators. Our main result in this paper is the recognition of the possible
convergence of the $\zeta$ series, and the
existence of some ``coupling constants'' that
determine the possible truncation of the $\zeta$ series at any desired order.
When this situation is encountered in a subclass of small-field
{\it slow-roll} inflationary models with canonical kinetic terms,
the one-loop corrections may dominate the series when calculating either the spectrum $P_\zeta$, or the bispectrum $B_\zeta$. This in turn {\it may generate sizeable and observable levels of non-gaussianity} in total contrast with the general claims found in the present literature.

The layout of the paper is the following:  in Section \ref{descriptors} we consider the quantities that describe the statistical properties encoded in any probability distribution function; theoretical explanations as well as observational constraints for $\zeta$ are given.  Section \ref{dNf} is devoted to the issue of the $\zeta$ series convergence and loop corrections in the framework of the $\delta N$ formalism, as well as to the presentation of the current knowledge about primordial non-gaussianity in slow-roll inflationary models.  A particular subclass of small-field slow-roll inflationary models is the subject of Section \ref{model} as it is this subclass of models that generate significant levels of non-gaussianity.  The available parameter space for this subclass of models is constrained in Section \ref{rest} by taking into account some observational requirements such as the COBE normalisation, the scalar spectral tilt, and the minimal amount of inflation. Another requirement of methodological nature,  the possible tree-level or one-loop dominance in $P_\zeta$ and/or $B_\zeta$, is considered in this section.  The level of non-gaussianity $f_{NL}$ in the bispectrum $B_\zeta$ is calculated in Section \ref{endcal} for models where $\zeta$ is 
generated during inflation; a comparison with the current literature is made.
Section \ref{seccou} is devoted to central issues in the consistency of the approach followed such as satisfying necessary conditions for the convergence of the $\zeta$ series and working in a perturbative regime.
Finally in Section \ref{concl} we conclude.
The professional reader who is already familiarized with the present ideas on the cosmological non-gaussianity may skip Sections \ref{descriptors} and \ref{dNf}, leaping directly to the new material starting from Section \ref{model}.
As regards the level of non-gaussianity $\tau_{NL}$ in the trispectrum $T_\zeta$, it will be studied following the sequence of ideas presented above in a companion paper \cite{cogollo}.

\section{Statistical descriptors for a probability distribution function} \label{descriptors}

The primordial curvature perturbation $\zeta$, as well as the contrast in the temperature of the cosmic microwave background radiation $\delta T/T$ and the gravitational potential $\Phi_g$, are examples of cosmological functions of space and time being described by probability distribution functions.  In particular, the probability distribution function $f(\zeta)$ for $\zeta$ has well defined statistical descriptors which depend directly upon the particular inflationary model and that are suitable for comparison with present observational data. Such a comparison allows us either to reject or to keep particular inflationary models as those which better represent nature's behaviour.  In this section we present a cosmologically motivated description of the statistical descriptors for probability distribution functions, focusing mainly on $f(\zeta)$. 

\subsection{Theoretical}
A probability distribution function $f(\zeta)$ for any function of space and time $\zeta({\bf x},t)$ may be understood as the univocal correspondence between the possible values that $\zeta$ may take throughout the space and the normalised frecuency of appearences of such values for a given time. Any continous function of $\zeta$ might represent a probability distribution function as long as $f(\zeta) \geq 0$ and $\int^\infty_{-\infty} f(\zeta) d\zeta = 1$. However, for a particular probability distribution function, how many independent parameters do we need to completely characterize it in a unique way? And despite the possible infinite number of parameters required to do this, what is the information encoded in those parameters? 
The answers to these questions rely on the moments $m_\zeta(n)$ of the distribution. 

For a given probability distribution function $f(\zeta)$, there are an infinite number of moments that work as statistical descriptors of $\zeta({\bf x},t)$:
\begin{eqnarray}
{\rm the \ mean \ value:} \;\; m_\zeta(1) &\equiv& \langle \zeta \rangle = \int \zeta f(\zeta) d\zeta \,, \\
{\rm the \ variance:} \;\; m_\zeta(2) &\equiv& \int (\zeta - \langle \zeta \rangle)^2 f(\zeta) d\zeta \,, \\
{\rm the \ skewness:} \;\; m_\zeta(3) &\equiv& \int (\zeta - \langle \zeta \rangle)^3 f(\zeta) d\zeta \,, \\
{\rm the \ kurtosis:} \;\; m_\zeta(4) &\equiv& \int (\zeta - \langle \zeta \rangle)^4 f(\zeta) d\zeta \,, \\
&.& \nonumber \\
&.& \nonumber \\
&.& \nonumber \\
&{\rm and \ so \ on.}& \nonumber
\end{eqnarray}

What can we say about $\zeta({\bf x},t)$ from the knowledge of the moments of the distribution? If, for instance, all the odd moments with $n \geq 3$ (skewness, ... etc) are zero, we can say that the probability distribution function $f(\zeta)$ is even around the mean value.  If in addition all the even moments with $n \geq 4$ (kurtosis, ... etc) are expressed only as products of the variance, we can say that the distribution function is gaussian. Indeed, as is well known, the only quantities required to reproduce a gaussian function are the mean value and the variance:
\begin{eqnarray}
f_{gaussian}(\zeta) \equiv \frac{1}{\sqrt{2\pi m_\zeta(2)}} e^{-(\zeta - m_\zeta(1))^2/2m_\zeta(2)} \,.
\end{eqnarray}
Departures from the exact gaussianity come either from non-vanishing odd moments with $n \geq 3$, in which case 
the probability distribution function is non-symmetric around the mean value, or from 
higher even moments different to products of the variance, in which case the probability distribution function continues to be symmetric around the mean value although it is non-gaussian, or from both 
of them.
A non-gaussian probability distribution function requires then more moments, other than the mean value and the variance, to be completely reconstructed. Such a reconstruction process is described for instance in Ref. \cite{sasaki1}.

Working in momentum space is especially useful in cosmology because the modes associated with the quantum fluctuations of scalar fields during inflation become classical once they leave the horizon \cite{liddle,lyth8,bernardeu3}. The same applies for the primordial curvature perturbation $\zeta$ which, in addition, is a conserved quantity while staying outside the horizon if the adiabatic condition is satisfied \cite{lyth4}. As regards the moments of the probability distribution function, they have a direct connection with the correlation functions for the Fourier modes $\zeta_{\bf k} = \int d^3k \zeta ({\bf x}) e^{-i {\bf k \cdot x}}$ defined in flat space.
As the $n$-point correlators of $\zeta_{\bf k}$ are generically defined in terms of spectral functions of the wavevectors involved\footnote{The homogeneity and isotropy requirements at large scales imply that the spectrum $P_\zeta$ and bispectrum $B_\zeta$ are functions of the wavenumbers only. For the trispectrum $T_\zeta$ and the other higher order spectral functions, the momentum dependence also involves the direction of the wavevectors.}:
\begin{eqnarray}
{\rm two-point \ correlation} &\rightarrow& {\rm spectrum} \ P_\zeta: \nonumber \\
\langle \zeta_{\bf k_1} \zeta_{\bf k_2} \rangle &\equiv& (2\pi)^3 \delta^3 ({\bf k_1} + {\bf k_2}) P_\zeta(k) \,, \label{2pc} \\
{\rm three-point \ correlation} &\rightarrow& {\rm bispectrum} \ B_\zeta: \nonumber \\
\langle \zeta_{\bf k_1} \zeta_{\bf k_2} \zeta_{\bf k_3} \rangle &\equiv& (2\pi)^3 \delta^3 ({\bf k_1} + {\bf k_2} + {\bf k_3}) B_\zeta(k_1, k_2, k_3) \,, \label{3pc} \\
{\rm four-point \ correlation} &\rightarrow& {\rm trispectrum} \ T_\zeta: \nonumber \\
\langle \zeta_{\bf k_1} \zeta_{\bf k_2} \zeta_{\bf k_3} \zeta_{\bf k_4} \rangle &\equiv& (2\pi)^3 \delta^3 ({\bf k_1} + {\bf k_2} + {\bf k_3} + {\bf k_4}) T_\zeta({\bf k_1}, {\bf k_2}, {\bf k_3}, {\bf k_4}) \,, \\
&.& \nonumber \\
&.& \nonumber \\
&.& \nonumber \\
&{\rm and \ so \ on,}& \nonumber
\end{eqnarray}
the moments of the distribution are then written as momentum integrals of the spectral functions for the modes $\zeta_{\bf k}$:
\begin{eqnarray}
{\rm the \ variance:} \ m_\zeta (2) &=& \int \frac{d^3 k}{(2\pi)^3} P_\zeta (k) \,, \\
{\rm the \ skewness:} \ m_\zeta (3) &=& \int \frac{d^3 k_1 \ d^3 k_2}{(2\pi)^6} B_\zeta (k_1,k_2,k_3) \,, \\
{\rm the \ kurtosis:} \ m_\zeta (4) &=& \int \frac{d^3 k_1 \ d^3 k_2 \ d^3 k_3}{(2\pi)^9} T_\zeta ({\bf k_1}, {\bf k_2}, {\bf k_3}, {\bf k_4}) \,, \\
&.& \nonumber \\
&.& \nonumber \\
&.& \nonumber \\
&{\rm and \ so \ on.}& \nonumber
\end{eqnarray}
Non-gaussianity in $\zeta$ is, therefore, associated with non-vanishing higher order spectral functions, starting from the bispectrum $B_\zeta$.

Theoretical cosmologists work with $\zeta$. However, astronomers work with observable quantities such as the contrast in the temperature of the cosmic microwave background radiation $\delta T/T$. The connection between the theoretical cosmologist quantity $\zeta$ and the astronomer quantity $\delta T/T$ is given by the Sachs-Wolfe effect \cite{sachs} which, at first-order and for superhorizon scales, looks as follows:
\begin{eqnarray}
\left(\frac{\delta T}{T}\right)_{\bf k} = - \frac{1}{5} \zeta_{\bf k} \,. \label{swe}
\end{eqnarray}
Thus, although it is essential to study the Sachs-Wolfe relation at higher orders, which is far more complicated than Eq. (\ref{swe}), theoretical cosmologists may study the statistical properties of the observed $\delta T / T$ through the spectral functions associated with the curvature perturbation $\zeta$:
\begin{eqnarray}
{\rm mean \ value \ of} \ \delta T/T = 0 &\rightarrow& {\rm mean \ value \ of} \ \zeta = 0 \,, \\
{\rm variance:} \;\; m_{\delta T/T}(2) &\rightarrow& {\rm spectrum:} \;\; P_\zeta (k) \,, \\
{\rm skewness:} \;\; m_{\delta T/T}(3) &\rightarrow& {\rm bispectrum:} \;\; B_\zeta (k_1,k_2,k_3) \,, \\
{\rm kurtosis:} \;\; m_{\delta T/T}(4) &\rightarrow& {\rm trispectrum:} \;\; T_\zeta ({\bf k_1}, {\bf k_2}, {\bf k_3}, {\bf k_4}) \,, \\
&.& \nonumber \\
&.& \nonumber \\
&.& \nonumber \\
&{\rm and \ so \ on.}& \nonumber
\end{eqnarray}

To end this section, we will parametrize the spectral functions of $\zeta$ in terms of quantities which are the ones for which observational bounds are given. Because of the direct connection between these quantities and the moments of the probability distribution function $f(\zeta)$, we may also call these quantities as the statistical descriptors for $f(\zeta)$. The spectrum $P_\zeta$ is parametrized in terms of an amplitude $\mathcal{P}^{1/2}_\zeta$ and a spectral index $n_\zeta$ which measures the deviation from an exactly scale-invariant spectrum \cite{liddle}:
\begin{equation}
P_\zeta (k) \equiv \frac{2\pi^2}{k^3} \mathcal{P}_\zeta \left(\frac{k}{aH}\right)^{n_\zeta - 1} \,, \label{asidef}
\end{equation}
where $a$ is the global expansion parameter and $H = \dot{a}/a$ is the Hubble parameter, with the dot meaning a derivative with respect to cosmic time.
The bispectrum $B_\zeta$ and trispectrum $T_\zeta$ are parametrized in terms of products of the spectrum $P_\zeta$, and the quantities $f_{NL}$ and $\tau_{NL}$ respectively\footnote{There is actually a sign difference between the $f_{NL}$ defined here and that defined in Ref. \cite{maldacena}.  The origin of the sign difference lies in the way the observed $f_{NL}$ is defined \cite{komatsu}, through the Bardeen's curvature perturbation: $\Phi^B = \Phi^B_L + f_{NL} (\Phi^B_L)^2$ with $\Phi^B = (3/5)\zeta$, and the way $f_{NL}$ is defined in Ref. \cite{maldacena}, through the gauge invariant Newtonian potential: $\Phi^N = \Phi^N_L + f_{NL} (\Phi^N_L)^2$ with $\Phi^N = -(3/5)\zeta$ \cite{com}.} \cite{boubekeur1,maldacena}:
\begin{eqnarray}
B_\zeta (k_1,k_2,k_3) &\equiv& \frac{6}{5} f_{NL} \left[P_\zeta(k_1) P_\zeta(k_2) + {\rm cyclic \ permutations} \right] \,, \label{bfp} \\
T_\zeta ({\bf k_1}, {\bf k_2}, {\bf k_3}, {\bf k_4}) &\equiv& \frac{1}{2} \tau_{NL} \left[P_\zeta(k_1) P_\zeta(k_2) P_\zeta(|{\bf k_1} + {\bf k_4}|) + {\rm cyclic \ permutations} \right] \,.
\end{eqnarray}
Higher order spectral functions would be parametrized in an analogous way.
Given the present observational state-of-the-art, $n_\zeta$, $f_{NL}$, and $\tau_{NL}$ are the statistical descriptors that discriminate among models for the origin of the large-scale structure once $\mathcal{P}^{1/2}_\zeta$ has been fixed to the observed value.  Since non-vanishing higher order spectral functions such as $B_\zeta$ and $T_\zeta$ imply non-gaussianity in the primordial curvature perturbation $\zeta$, the statistical descriptors $f_{NL}$ and $\tau_{NL}$ are usually called the levels of non-gaussianity. 

\subsection{Observational} \label{observational}

COBE provided us with a reliable value for the spectral amplitude $\mathcal{P}^{1/2}_\zeta$ \cite{bunn}: $\mathcal{P}^{1/2}_\zeta = (4.957 \pm 0.094) \times 10^{-5}$ which is usually called the COBE normalisation.  As regards the spectral index, the latest data release and analysis from the WMAP satellite shows that $n_\zeta = 0.960 \pm 0.014$ \cite{komatsu1} which rejects exact scale invariance at more than $2\sigma$. Such a result has been extensively used to constrain inflation model building \cite{alabidi3}, and although several classes of inflationary models have been ruled out through the spectral index, lots of models are still allowed; that is why it is so important an appropiate knowledge of the statistical descriptors $f_{NL}$ and $\tau_{NL}$.
Present observations show that the primordial curvature perturbation $\zeta$ is almost, but not completely, gaussian.  The level of non-gaussianity $f_{NL}$ in the bispectrum $B_\zeta$, 
after five years of data from NASA's WMAP satellite, is in the range $-9 < f_{NL} < 111$ at $2\sigma$ \cite{komatsu1}. There is at present no observational bound on the level of non-gaussianity $\tau_{NL}$ in the trispectrum $T_\zeta$ although it was predicted that COBE should either detect it or impose the lower bound $|\tau_{NL}| \lsim 10^8$ \cite{boubekeur1,okamoto}. It is expected that future WMAP data releases will either detect non-gaussianity or reduce the bounds on $f_{NL}$ and $\tau_{NL}$ at the $2\sigma$ level to $|f_{NL}|\lsim 40$ \cite{komatsu} and $|\tau_{NL}|\lsim
2 \times 10^{4}$ \cite{kogo} respectively. The ESA's PLANCK satellite \cite{planck,planck1}, which will be launched in October 2008, promises to reduce the bounds to $|f_{NL}|\lsim 10$ \cite{komatsu} and $|\tau_{NL}|\lsim 560$  \cite{kogo} at the $2\sigma$ level if non-gaussianity is not detected. In addition, by studying the 21-cm emission spectral line in the cosmic neutral Hydrogen prior to the era of reionization, it is also possible to know about the levels of non-gaussianity $f_{NL}$ and $\tau_{NL}$;
the 21-cm background anisotropies capture information about the primordial non-gaussianity better than any high resolution map of cosmic microwave background radiation: an experiment like this could reduce the bounds on the non-gaussianity levels to $|f_{NL}|\lsim 0.2$ \cite{cooray1,cooray2} and $|\tau_{NL}|\lsim 20$ \cite{cooray2} at the $2\sigma$ confidence. Finally, it is worth stating that there have been recent claims about the detection of non-gaussianity in the bispectrum $B_\zeta$ of $\zeta$ from the WMAP 3-year data \cite{yadav,jeong}. Such claims, which report a rejection of $f_{NL} = 0$ at more that $2\sigma$ ($26.9 < f_{NL} < 146.7$), are based on the estimation of the bispectrum while using some specific foreground masks.  The WMAP 5-year analysis \cite{komatsu1} shows a similar behaviour when using those masks, but reduces the significance of the results when other more conservative masks are included allowing again the possibility of exact gaussianity.

\section{The $\delta N$ formalism} \label{dNf}

The $\delta N$ formalism \cite{starobinsky,sasaki2,lyth4} provides a powerful method for calculating $\zeta$ and all its statistical descriptors at any desired order in cosmological perturbation theory from knowing, in the case of slow-roll inflationary models, only the evolution of a family of unperturbed universes and the correlation functions a bit later than horizon exit of the perturbations in the field scalars present during inflation \cite{lyth2}. This section is devoted to a brief review of the formalism and to a short discussion of some relevant issues which either have not been properly taken into account or have not been discussed at all in the current literature.  

\subsection{The non-linear primordial curvature perturbation $\zeta$}

In the cosmological standard model \cite{mukhanov} the observable Universe is homogeneous and isotropic, being described by the unperturbed Friedmann-Robertson-Walker metric whose line element, for a spatially flat universe, looks as follows:
\begin{equation}
ds^2 = -dt^2 + a^2(t) \delta_{ij} dx^i dx^j \,,
\end{equation}
where $a(t)$ is the global expansion parameter, $t$ is the cosmic time, and ${\bf x}$ represents the position in cartesian spatial coordinates. The homogeneity and isotropy conditions describe very well the Universe at large scales, but departures from the unperturbed background are observationally evident at smaller scales.

One way to parametrize the departures from the homogeneous and isotropic background is to include perturbations in the metric, for which we have to define a slicing and a threading. The slicing will be defined so that the energy density in fixed-$t$ slices of spacetime is uniform. The threading will correspond to comoving fixed-$x$ world lines. Thus, the perturbed spatial metric may be defined as
\begin{equation}
g_{ij} \equiv a^2(t)e^{2\zeta(t,{\bf x})} \gamma_{ij}(t,{\bf x}) \,, \label{spatialm}
\end{equation}
where $\gamma_{ij} (t,{\bf x})$, which gives origin to the tensor perturbations, has unit determinant. This means that we may define a local scale factor
\begin{equation}
\tilde{a} (t,{\bf x}) = a(t) e^{\zeta(t,{\bf x})} \,, \label{localep}
\end{equation}
which is proportial to any volume of the Universe bounded by fixed spatial coordinates.

The interesing feature of Eq. (\ref{localep}) is that it is possible to write the $\zeta (t,{\bf x})$ quantity in terms of the perturbation in the amount of expansion from an initial time $t_{\rm in}$ where the slice is flat (i.e. the spatial metric in such a slice is the same as that in Eq. (\ref{spatialm}) but without the $\zeta$ factor) to a final time $t$ where the slice is of uniform energy density:
\begin{eqnarray}
\zeta(t,{\bf x}) \equiv \delta N &=& N(t,{\bf x}) - N_0 (t) \label{mdn} \\
&=& \ln \left[\frac{\tilde{a}(t,{\bf x})}{a(t_{\rm in})}\right] - \ln \left[\frac{a(t)}{a(t_{\rm in})}\right] \,.
\end{eqnarray}
This is the $\delta N$ formalism \cite{starobinsky,sasaki2,lyth4} where the $\zeta(t,{\bf x})$ quantity, being a non-linear extension of the primordial curvature pertubation, reproduces the usually defined gauge-invariant curvature perturbation $\zeta_1$ at first order \cite{bardeen}:
\begin{equation}
\zeta_1 = -\psi - H \frac{\delta_1 \rho}{\dot{\rho}} \,,
\end{equation}
where $\rho$ is the energy density 
and $\psi$ is the scalar perturbation in the spatial metric at superhorizon scales and at first order:
\begin{equation}
g_{ij} = a^2(t) (1 - 2\psi) \delta_{ij} \,.
\end{equation}
$\zeta$, as non-linearly defined in Eqs. (\ref{spatialm}) and (\ref{mdn}), is a conserved quantity on superhorizon scales as long as the adiabatic condition (the pressure as a function of only the energy density) is satisfied \cite{lyth4}.

\subsection{$\zeta$ series convergence and loop corrections}

In order to calculate $\zeta (t,{\bf x})$ from Eq. (\ref{mdn}), we need information about the physical content of the Universe at times $t$ and $t_{\rm in}$. By choosing the initial time $t_{\rm in}$ a few Hubble times after the cosmologically relevant scales leave the horizon during inflation $t_{\rm in} = t_\star$, and the final time $t$ corresponding to a slice of uniform energy density, we recognize that $N$, for slow-roll inflationary models, is completely parametrized by the values a few Hubble times after horizon exit of the scalar fields $\phi_i$ present during inflation and the energy density at the time at which one wishes to calculate $\zeta$:
\begin{equation}
\zeta (t,{\bf x}) \equiv N(\rho(t),\phi_1(t_\star,{\bf x}),\phi_2(t_\star,{\bf x}), ...) - N(\rho(t),\phi_1(t_\star),\phi_2(t_\star), ...) \,.
\end{equation}
The previous expression can be Taylor-expanded around the unperturbed background values for the scalar fields $\phi_i$ and suitably redefined so that $\langle \zeta(t,{\bf x}) \rangle = 0$. Thus, 
\begin{eqnarray}
\zeta(t,\textbf{x})&=&\sum_{i}\textit{N}_{i}(t)\delta\phi_{i}(t_\star,{\bf x}) - \sum_{i}\textit{N}_{i}(t) \langle \delta\phi_{i}(t_\star,{\bf x}) \rangle + \nonumber \\
&& + \frac{1}{2}\sum_{ij}\textit{N}_{ij}(t)\delta\phi_{i}(t_\star,{\bf x})\delta\phi_{j}(t_\star,{\bf x}) - \frac{1}{2}\sum_{ij}\textit{N}_{ij}(t) \langle \delta\phi_{i}(t_\star,{\bf x})\delta\phi_{j}(t_\star,{\bf x}) \rangle + \nonumber \\
&& + \frac{1}{3!}\sum_{ijk}\textit{N}_{ijk}(t)\delta\phi_{i}(t_\star,{\bf x})\delta\phi_{j}(t_\star,{\bf x})
\delta\phi_{k}(t_\star,{\bf x}) - \frac{1}{3!}\sum_{ijk}\textit{N}_{ijk}(t) \langle \delta\phi_{i}(t_\star,{\bf x})\delta\phi_{j}(t_\star,{\bf x})
\delta\phi_{k}(t_\star,{\bf x}) \rangle + \nonumber \\
&& + ...\;, \label{Nexp}
\end{eqnarray}
where the $\delta \phi_i (t_\star,{\bf x})$ are the scalar field perturbations in the flat slice a few Hubble times after horizon exit, whose spectrum amplitude is given by \cite{bunch}
\begin{equation}
\mathcal{P}^{1/2}_{\delta \phi_{i}} = 
\frac{H_\star}{2\pi} 
\,,
\end{equation}
and the notation for the $N$ derivatives is $N_i\equiv\frac{\partial N}{\partial\phi_{i}}$,
$N_{ij}\equiv\frac{\partial^{2}N}{\partial\phi_{i}\partial\phi_{j}}$, and so on.

The expression in Eq. (\ref{Nexp}) has been used to calculate the statistical descriptors of $\zeta$ at any desired order in cosmological perturbation theory by consistently truncating the series \cite{lyth2}.  For instance, by truncating the series at first order, the amplitude of the spectrum $P_\zeta$ of $\zeta$ defined in Eqs. (\ref{2pc}) and (\ref{asidef}) is given by \cite{sasaki2}
\begin{equation}
\mathcal{P}_\zeta = \left(\frac{H_\star}{2\pi}\right)^2 \sum_i N_i^2 \,, \label{adnf}
\end{equation}
which in turn gives the well known formula for the spectral index \cite{sasaki2}:
\begin{equation}
n_\zeta - 1 = -2\epsilon - 2m_P^2 \frac{\sum_{ij} V_i N_j N_{ij}}{V \sum_i N_i^2} \,, \label{ndnf}
\end{equation}
where a subindex $i$ in $V$ means a derivative with respect to the $\phi_i$ field, and
being $\epsilon$ one of the slow-roll parameters defined by $\epsilon = -\dot{H}/H^2$, $m_P = (8\pi G)^{-2}$ the reduced Planck mass, and $V$ the scalar inflationary potential.
Analogously, the level of non-gaussianity $f_{NL}$ in the bispectrum $B_\zeta$ of $\zeta$ defined in Eqs. (\ref{3pc}) and (\ref{bfp}) is obtained by truncating the series at second order and assuming that the scalar field perturbations $\delta \phi_i$ are perfectly gaussian \cite{lyth2}:
\begin{equation}
\frac{6}{5} f_{NL} = \frac{\sum_{ij} N_i N_j N_{ij}}{\left[\sum_i N_i^2\right]^2} +  \mathcal{P}_\zeta \frac{\sum_{ijk} N_{ij} N_{jk} N_{ki}}{\left[\sum_i N_i^2\right]^3} \ln(kL) \,. \label{fdnf}
\end{equation}
In the last expression the $\ln(kL)$ factor is of order one, $L$ being the infrared cutoff when calculating the stochastic properties in a minimal box \cite{lyth1,bernardeu4}.

The truncated series methodology has proved to be powerful and reliable at reproducing successfully the level of non-gaussianity $f_{NL}$ in single-field slow-roll models \cite{seery5} and in the curvaton scenario \cite{lyth2}. Nevertheless, for more general models, how reliable is it to truncate the series at some order? In the first place, from Eq. (\ref{Nexp}) it is impossible to know whether the series converges until the $N$ derivatives are explicitly calculated and the convergence radius is obtained;  obviously if the series is not convergent at all, the expansion in Eq. (\ref{Nexp}) is meaningless.  Without any proof of the contrary, the current assumption in the literature \cite{sasaki2,lyth2,boubekeur1,alabidi2,zaballa,alabidi1,vernizzi,battefeld,yokoyama1,yokoyama2,seery3,byrnes2} has been that the $\zeta$ series is convergent. In addition,
supposing that the convergence radius is finally known,
the truncation at any desired order would again be meaningless if some leading terms in the series get excluded.  Such a situation might easily happen if each ${\bf x}$-dependent term in the $\zeta$ series is considered smaller than the previous one, which indeed is the standard assumption \cite{sasaki2,lyth2,boubekeur1,alabidi2,zaballa,alabidi1,vernizzi,battefeld,yokoyama1,yokoyama2,seery3,byrnes2}, but which is not a universal fact.

When studying the series through a diagrammatic approach \cite{byrnes1}, in an analogous way to that for Quantum Field Theory via Feynman diagrams, the first-order terms in the spectral functions are called the tree-level terms. Examples of these tree-level terms are those in Eqs. (\ref{adnf}) and (\ref{ndnf}), and the first one in Eq. (\ref{fdnf}). Higher-order corrections, such as that which contributes with the second term in Eq. (\ref{fdnf}), are called the loop terms because they involve internal momentum integrations. The statistical descriptors of $\zeta$ has been so far studied by naively neglecting the loop corrections against the tree-level terms \cite{sasaki2,lyth2,boubekeur1,alabidi2,zaballa,alabidi1,vernizzi,battefeld,yokoyama1,yokoyama2,seery3,byrnes2}; nevertheless, as might happen in Quantum Field Theory, eventually some loop corrections could be bigger than the tree-level terms, so it is essential to properly study the possible $n$-loop dominance in the spectral functions.

\subsection{Non-gaussianity in slow-roll inflation}

The most frequent class of inflationary models found in the literature are those which satisfy the so called slow-roll conditions, as these very simple models easily meet the spectral index observational requirements discussed in Subsection \ref{observational} for the generation of large-scale structures.

The slow-roll conditions for single-field inflationary models with canonical kinetic terms read
\begin{eqnarray}
\dot{\phi}^2 &\ll& V(\phi) \,, \label{1stsrc} \\
|\ddot{\phi}| &\ll& |3H\dot{\phi}| \,, \label{2ndsrc}
\end{eqnarray}
where 
$\phi$ is the inflaton field and $V(\phi)$ is the scalar field potential. 
On defining the slow-roll parameters $\epsilon$ and $\eta_\phi$ as \cite{liddle}
\begin{eqnarray}
\epsilon &\equiv& -\frac{\dot{H}}{H^2} \,, \label{1stsrp} \\
\eta_\phi &\equiv& \epsilon - \frac{\ddot{\phi}}{H\dot{\phi}} \,, \label{2ndsrp}
\end{eqnarray}
the slow-roll conditions in Eqs. (\ref{1stsrc}) and (\ref{2ndsrc}) translate into strong constraints for the slow-roll parameters: $\epsilon, |\eta_\phi| \ll 1$, which actually become $\epsilon, |\eta_\phi| \lsim 10^{-2}$ in view of Eq. (\ref{ndnf}) for single-field inflation:
\begin{equation}
n_\zeta - 1 = 2\eta_\phi - 6\epsilon \,,
\end{equation}
and the observational requirements presented in Subsection \ref{observational}.

Multifield slow-roll models may also be characterized by a set of slow-roll parameters which generalize those in Eqs. (\ref{1stsrp}) and (\ref{2ndsrp}) \cite{lyth5}:
\begin{eqnarray}
\epsilon_i &\equiv& \frac{m_P^2}{2} \left(\frac{V_i}{V}\right)^2 \,, \\
\eta_i &\equiv& m_P^2 \frac{V_{ii}}{V} \,.
\end{eqnarray}
By writing the slow-roll parameters in terms of derivatives of the scalar potential, as in the last two expressions, we realize that the slow-roll conditions require very flat potentials to be met.

The level of non-gaussianity $f_{NL}$ in slow-roll inflationary models with canonical kinetic terms has been studied both for single-field \cite{maldacena} and for multiple-field inflation \cite{vernizzi,battefeld,yokoyama1}, assuming $\zeta$ series convergence and considering only the tree-level terms both in $P_\zeta$ and in $B_\zeta$.  What these works find is a strong dependence on the slow-roll parameters $\epsilon_i$ and $\eta_i$; for instance, Ref. \cite{maldacena} gives us for single-field models:
\begin{equation}
\frac{6}{5} f_{NL} = \epsilon(1 + f) + 2\epsilon - \eta_\phi \,,
\end{equation}
where $f$ is a function of the shape of the wavevectors triangle within the range $0 \leq f \leq 5/6$. Refs. \cite{lyth1,bartolo} show that in such a case the inclusion of loop corrections is unnecessary because the latter are so small compared to the tree-level terms. Thus, $f_{NL}$ in single-field models with canonical kinetic terms is slow-roll suppressed and, therefore, unobservably small.  As regards the multifield models, $f_{NL}$ was shown, first in the case of two-field inflation with separable potential \cite{vernizzi} and later for multiple-field inflation with separable \cite{battefeld} and non-separable \cite{yokoyama1} potentials, to be a rather complex function of the slow-roll parameters and the scalar potential that in most of the cases ends up being slow-roll suppressed. Only for models with a sharply curved trajectory in field space might the $f_{NL}$ be at most of order one, the only possible examples to date being the models of Refs. \cite{alabidi1,bernardeu1,bernardeu2} 
(see anyway footnote \ref{laila}). Again, such predictions are based on the assumptions that the $\zeta$ series is convergent and that the tree-level terms are the leading ones, so they might be badly violated if loop corrections are considered.

Following a treatment parallel to that in Ref. \cite{vernizzi}, the level of non-gaussianity $\tau_{NL}$ is calculated in Ref. \cite{seery3} for multifield slow-roll inflationary models with canonical kinetic terms, separable potential, and assuming convergence of the $\zeta$ series and tree-level dominance.  From reaching similar conclusions to those found for the $f_{NL}$ case, the $\tau_{NL}$ is slow-roll suppressed in most of the cases although it might be of order one if the trajectory in field space is sharply curved. Nevertheless, as will be shown in a companion paper \cite{cogollo}, there may be a big enhancement in $\tau_{NL}$ if loop corrections are taken into account.

Finally, it is worth mentioning that there are other classes of models where the levels of non-gaussianity $f_{NL}$ and $\tau_{NL}$ are big enough to be observable.  Some of these models correspond to general langrangians with non-canonical kinetic terms ($k$-inflation \cite{armendariz}, DBI inflation \cite{silverstein}, ghost inflation \cite{arkani}, etc.), where the sizeable levels of non-gaussianity have mostly a quantum origin, i.e. their origin relies on the quantum correlators of the field perturbations a few Hubble times after horizon exit. Non-gaussianity in $B_\zeta$ has been studied in these models for the single-field case \cite{seery6,chen} and also for the multifield case \cite{gao,langlois1,langlois2,arroja2}. A recent paper discusses the non-gaussianity in $T_\zeta$ for these general models for single-field inflation \cite{arroja}. In contrast, there are some other models where the large non-gaussianities have their origin in the field dynamics at the end of inflation \cite{lyth9,bernardeu5}; nice examples of this proposal are studied for instance in Refs. \cite{matsuda1,matsuda2,sasakin1,sasakin2}. However, since the inflationary models of the slow-roll variety with canonical kinetic terms are the simplest, the most popular, and the best studied so far although, in principle, the non-gaussianity statistical descriptors are too small to ever be observable, it is very interesting to consider the possibility of having an example of such models which does generate {\it sizeable and observable values for} $f_{NL}$. This appealing possibility will be the subject of the following sections.

\section{A subclass of small-field slow-roll inflationary models} \label{model}

According to the classification of inflationary models proposed in Ref. \cite{dodelson2}, the small-field models are those of the form that would be expected as a result of spontaneous symmetry breaking, with a field initially near an unstable equilibrium point (usually taken to be at the origin) and rolling toward a stable minimum $\langle \phi \rangle \neq 0$.  Thus, inflation occurs when the field is small relative to its expectation value $\phi \ll \langle \phi \rangle$.  Some interesting examples are the original models of new inflation \cite{linde,albrecht}, modular inflation from string theory \cite{dimopoulos}, natural inflation \cite{freese}, and hilltop inflation \cite{boubekeur2}. As a result, the inflationary potential for small-field models may be taken as
\begin{equation}
V = \sum_i \Lambda_i \left[1 - \left(\frac{\phi_i}{\mu_i}\right)^p \right] \,,
\end{equation}
where the subscript $i$ here denotes the relevant quantities of the $i$th field, $p$ is the same for all fields, and $\Lambda_i$ and $\mu_i$ are the parameters describing the height and tilt of the potential of the $i$th field.

\begin{figure*} [t]
\begin{center}
\includegraphics[width=10cm,height=15cm,angle=-90]{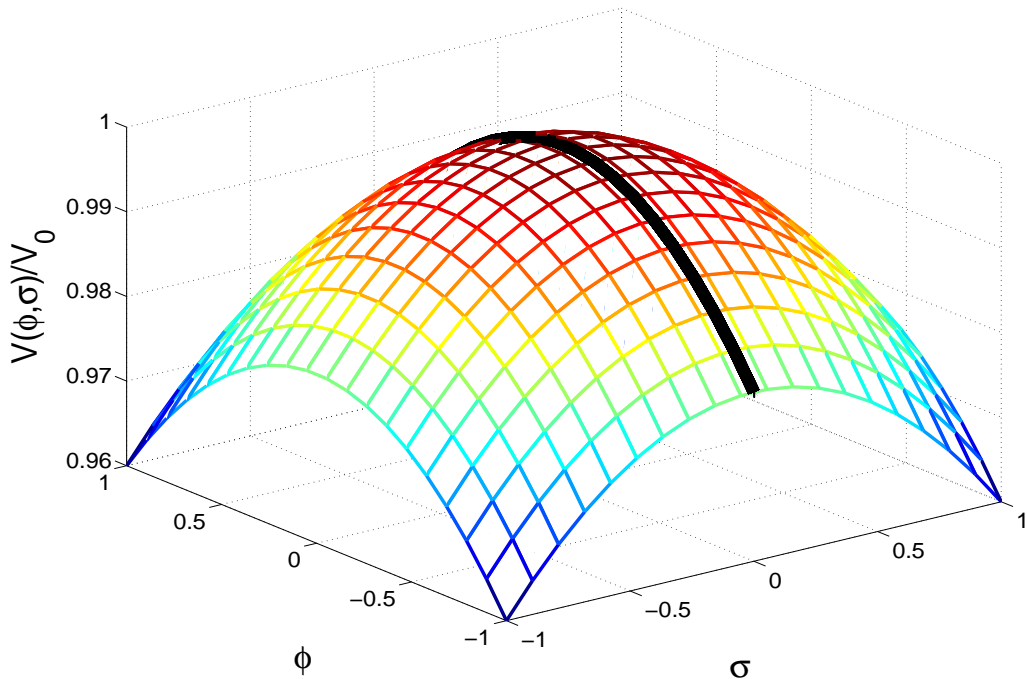}
\end{center}
\caption{Our small-field slow-roll potential of Eq. (\ref{pot}) with $\eta_\phi,\eta_\sigma < 0$. The inflaton starts near the maximum and moves away from the origin following the $\sigma = 0$ trajectory depicted with the solid black line. (This figure has been taken from Ref. \cite{alabidi1}).}
\label{fig3}
\end{figure*}

While Ref. \cite{ahmad} studies the spectrum of $\zeta$ for general values of the parameter $p$ and an arbitrary number of fields, assuming $\zeta$ series convergence and tree-level dominance, we will specialize to the $p=2$ case for two fields $\phi$ and $\sigma$:
\begin{equation}
V = V_0\left(1+\frac{1}{2}\eta_\phi \frac{\phi^2}{m_P^2} +\frac{1}{2}\eta_\sigma \frac{\sigma^2}{m_P^2}\right) \,, \label{pot}
\end{equation}
where we have traded the expressions
\begin{eqnarray}
\Lambda_1 + \Lambda_2 &{\rm for}& V_0 \,, \\
\frac{\Lambda_1}{\mu_1^2} &{\rm for}& -V_0 \frac{\eta_\phi}{2m_P^2} \,,
\end{eqnarray}
and
\begin{equation}
\frac{\Lambda_2}{\mu_2^2} \; \; \; {\rm for} \; \; \; -V_0 \frac{\eta_\sigma}{2m_P^2} \,.
\end{equation}
On doing this, and assuming that the first term in Eq. (\ref{pot}) dominates, $\eta_\phi < 0$ and $\eta_\sigma < 0$ become the usual $\eta$ slow-roll parameters associated with the fields $\phi$ and $\sigma$.

We have chosen
for simplicity the $\sigma=0$ trajectory (see Fig. \ref{fig3}) since in that case the potential in Eq. (\ref{pot}) reproduces for some number of e-folds the hybrid inflation scenario \cite{linde2} where $\phi$ is the inflaton and $\sigma$ is the waterfall field. Non-gaussianity in such a model has been studied in Refs. \cite{lyth2,zaballa,alabidi1,lyth3,enqvist1,vaihkonen}; in particular, Ref. \cite{lyth2} used a one-loop correction to conjecture that $f_{NL}$ in this model would be sizeable {\it only if} $\zeta$ {\it was not} generated during inflation, which turns out not to be a necessary requirement as we will show later. Ref. \cite{alabidi1}, in contrast, works only at tree-level with the same potential as Eq. (\ref{pot}) but relaxing the $\sigma = 0$ condition, finding that values for $f_{NL} \sim \mathcal{O}(1)$ are possible for a small set of initial conditions and assuming a saddle-point like form for the potential ($\eta_\phi < 0$ and $\eta_\sigma > 0$).

\section{Constraints for having a reliable parameter space} \label{rest}

We will explore now the constraints that the model must satisfy before we calculate $f_{NL}$. Our guiding idea will be the consideration of the role that the tree-level terms and one-loop corrections to both $P_\zeta$ and $B_\zeta$ have in the determination of the available parameter space.  Only after calculating $f_{NL}$ in Section \ref{endcal} will we come back to the discussion of the consistency of the approach followed in the present section by studying the $\zeta$ series convergence and the validity of the truncation at one loop level.

\subsection{Tree-level or one-loop dominance} \label{secreg}
Since we are considering a slow-roll regime, the evolution of the background $\phi$ and $\sigma$ fields in such a case is given by the Klein-Gordon equation
\begin{equation}
\ddot{\phi} + 3H\dot{\phi} + V_\phi = 0 \,, 
\end{equation}
supplemented with the slow-roll condition in Eq. (\ref{2ndsrc}).  This leads to
\begin{eqnarray}
\phi(N) &=& \phi_\star \exp(-N\eta_\phi) \,, \label{srp} \\
\sigma(N) &=& \sigma_\star \exp(-N\eta_\sigma) \,,
\end{eqnarray}
so the potential above leads to the following derivatives of $N$ with respect to $\phi_\star$ and $\sigma_\star$ for the $\sigma = 0$ trajectory:\footnote{When calculating the $N$-derivatives, we have considered that the final time corresponds to a slice of uniform energy density. This means, in the slow-roll approximation, that $V$ is homogeneous.}
\begin{eqnarray}
&&N_\phi = \frac{1}{\eta_\phi \phi_\star} \,, \hspace{5mm}  N_\sigma = 0 \,, \label{1d} \\
&&N_{\phi \phi} = -\frac{1}{\eta_\phi \phi_\star^2} \,, \hspace{5mm} N_{\phi \sigma} = 0 \,, \hspace{5mm} N_{\sigma \sigma} = \frac{\eta_\sigma}{\eta_\phi^2 \phi_\star^2} \exp[2N (\eta_\phi - \eta_\sigma)] \,, \label{2d} \\
&&N_{\phi \phi \phi} = \frac{2}{\eta_\phi \phi_\star^3}\,, \hspace{5mm} N_{\phi \phi \sigma} = 0\,, \hspace{5mm} N_{\sigma \sigma \phi} = -\frac{2\eta_\sigma^2}{\eta_\phi^3 \phi_\star^3} \exp[2N (\eta_\phi - \eta_\sigma)]\,, \hspace{5mm} N_{\sigma \sigma \sigma} = 0\,, \label{3d} \\
&&... \nonumber \\
&&{\rm and \ so \ on}. \nonumber
\end{eqnarray}

By means of the $\delta N$ formalism, we can make use of the above formulae to calculate the spectrum and the bispectrum of the curvature perturbation including the tree-level and the one-loop contributions when $|\eta_\sigma| > |\eta_\phi|$ (see Appendix \ref{app}).  This is the interesting case since, as will be shown in Section \ref{endcal}, it generates sizeable values for $f_{NL}$.  Following the results in Appendix \ref{app},  we will write down just the leading terms to the tree-level and one-loop contributions given in Eqs. (\ref{pt1}), (\ref{1lfpd}), (\ref{tlbd1}), and (\ref{1lfbd}):
\begin{eqnarray}
\mathcal{P}_\zeta^{tree} &=& \frac{1}{\eta_\phi^2 \phi_\star^2} \left(\frac{H_\star}{2\pi}\right)^2 \,, \label{pt} \\
\mathcal{P}_\zeta^{1-loop} &=& \frac{\eta_\sigma^2}{\eta_\phi^4 \phi_\star^4} \exp[4N(|\eta_\sigma| - |\eta_\phi|)] \left(\frac{H_\star}{2\pi}\right)^4 \ln(kL) \,, \label{pl} \\
B_\zeta^{tree} &=& -\frac{1}{\eta_\phi^3 \phi_\star^4} \left(\frac{H_\star}{2\pi}\right)^4 4\pi^4 \left(\frac{\sum_i k_i^3}{\prod_i k_i^3}\right) \,, \label{bt} \\
B_\zeta^{1-loop} &=& \frac{\eta_\sigma^3}{\eta_\phi^6 \phi_\star^6} \exp[6N(|\eta_\sigma| - |\eta_\phi|)] \left(\frac{H_\star}{2\pi}\right)^6 \ln(kL) 4\pi^4 \left(\frac{\sum_i k_i^3}{\prod_i k_i^3}\right) \,. \label{bl}
\end{eqnarray}

Because of the exponential factors in Eqs. (\ref{pl}) and (\ref{bl}), it might be possible that the one-loop corrections dominate over $P_\zeta$ and/or $B_\zeta$.  There are three posibilities in complete connection with the position of the $\phi$ field when the cosmologically relevant scales are exiting the horizon:

\subsubsection{Both $B_\zeta$ and $P_\zeta$ are dominated by the one-loop corrections}
Comparing Eqs. (\ref{pt}) with (\ref{pl}) and Eqs. (\ref{bt}) with (\ref{bl}) we require in this case that
\begin{eqnarray}
\frac{\eta_\sigma^2}{\eta_\phi^2} \exp[4N(|\eta_\sigma|-|\eta_\phi|)] &\gg& \frac{1}{\frac{1}{\phi_\star^2} \left(\frac{H_\star}{2\pi}\right)^2} \,, \\
\frac{\eta_\sigma^3}{\eta_\phi^3} \exp[6N(|\eta_\sigma|-|\eta_\phi|)] &\gg& \frac{1}{\frac{1}{\phi_\star^2} \left(\frac{H_\star}{2\pi}\right)^2} \,,
\end{eqnarray}
in which case only the first inequality is required. Employing the definition for the tensor to scalar ratio $r$ \cite{lyth6}:
\begin{equation}
r \equiv \frac{\mathcal{P}_T}{\mathcal{P}_\zeta} = \frac{\frac{8}{m_P^2} \left(\frac{H_\star}{2\pi}\right)^2}{\mathcal{P}_\zeta} \,, \label{defr}
\end{equation}
$\mathcal{P}_T^{1/2}$ being the amplitude of the spectrum for primordial gravitational waves,
we can write such an inequality as
\begin{equation}
\left(\frac{\phi_\star}{m_P}\right)^2 \ll \frac{r \mathcal{P}_\zeta}{8} \frac{\eta_\sigma^2}{\eta_\phi^2} \exp[4N(|\eta_\sigma|-|\eta_\phi|)] \,.
\label{lowphi}
\end{equation}
From now on we will name the parameter window described by Eq. (\ref{lowphi}) as the low $\phi_\star$ region since the latter represents a region of allowed values for $\phi_\star$ limited by an upper bound.

\subsubsection{$B_\zeta$ dominated by the one-loop correction and $P_\zeta$ dominated by the tree-level term} \label{bz1l}
Comparing Eqs. (\ref{pt}) with (\ref{pl}) and Eqs. (\ref{bt}) with (\ref{bl}) we require in this case that
\begin{eqnarray}
\frac{\eta_\sigma^2}{\eta_\phi^2} \exp[4N(|\eta_\sigma|-|\eta_\phi|)] &\ll& \frac{1}{\frac{1}{\phi_\star^2} \left(\frac{H_\star}{2\pi}\right)^2} \,, \label{wn} \\
\frac{\eta_\sigma^3}{\eta_\phi^3} \exp[6N(|\eta_\sigma|-|\eta_\phi|)] &\gg& \frac{1}{\frac{1}{\phi_\star^2} \left(\frac{H_\star}{2\pi}\right)^2} \,, 
\end{eqnarray}
which combines to give, employing the definition for the tensor to scalar ratio $r$ introduced in Eq. (\ref{defr}),
\begin{equation}
\frac{r \mathcal{P}_\zeta}{8} \frac{\eta_\sigma^2}{\eta_\phi^2} \exp[4N(|\eta_\sigma|-|\eta_\phi|)] \ll \left(\frac{\phi_\star}{m_P}\right)^2 \ll \frac{r \mathcal{P}_\zeta}{8} \frac{\eta_\sigma^3}{\eta_\phi^3} \exp[6N(|\eta_\sigma|-|\eta_\phi|)] \,. 
\label{intc}
\end{equation}
From now on we will name the parameter window described by Eq. (\ref{intc}) as the intermediate $\phi_\star$ region since the latter represents a region of allowed values for $\phi_\star$ limited by both an upper and a lower bound.

\subsubsection{Both $B_\zeta$ and $P_\zeta$ are dominated by the tree-level terms}
Comparing Eqs. (\ref{pt}) with (\ref{pl}) and Eqs. (\ref{bt}) with (\ref{bl}) we require in this case that
\begin{eqnarray}
\frac{\eta_\sigma^2}{\eta_\phi^2} \exp[4N(|\eta_\sigma|-|\eta_\phi|)] &\ll& \frac{1}{\frac{1}{\phi_\star^2} \left(\frac{H_\star}{2\pi}\right)^2} \,, \\
\frac{\eta_\sigma^3}{\eta_\phi^3} \exp[6N(|\eta_\sigma|-|\eta_\phi|)] &\ll& \frac{1}{\frac{1}{\phi_\star^2} \left(\frac{H_\star}{2\pi}\right)^2} \,,
\end{eqnarray}
in which case only the second inequality is required.  Employing the definition for the tensor to scalar ratio $r$ introduced in Eq. (\ref{defr}), we can write such an inequality as
\begin{eqnarray}
\left(\frac{\phi_\star}{m_P}\right)^2 \gg \frac{r \mathcal{P}_\zeta}{8} \frac{\eta_\sigma^3}{\eta_\phi^3} \exp[6N(|\eta_\sigma|-|\eta_\phi|)] \,. 
\label{highphi}
\end{eqnarray}
From now on we will name the parameter window described by Eq. (\ref{highphi}) as the high $\phi_\star$ region, since the latter represents a region of allowed values for $\phi_\star$ limited by a lower bound. 

\subsection{Spectrum normalisation condition} \label{secnorm}
The model must satisfy the COBE normalisation on the spectrum amplitude $\mathcal{P}^{1/2}_\zeta$ \cite{bunn} considering that $\zeta$ is assumed in this paper to be generated during inflation\footnote{The scenario where $\zeta$ is assumed not to be generated during inflation will be presented in a companion paper \cite{cogollo}.}.
There exist two possibilities discussed right below.

\subsubsection{$\zeta$ generated during inflation and $P_\zeta$ dominated by the one-loop correction}
According to Eq. (\ref{pl}), and the tensor to scalar ratio $r$ definition in Eq. (\ref{defr}), we have in this case
\begin{eqnarray}
\mathcal{P}_\zeta^{1-loop} &=& \frac{\eta_\sigma^2}{\eta_\phi^4 \phi_\star^4} \exp[4N(|\eta_\sigma| - |\eta_\phi|)] \left(\frac{H_\star}{2\pi}\right)^4 \ln(kL) \nonumber \\
&=& \frac{\eta_\sigma^2}{\eta_\phi^4} \exp[4N(|\eta_\sigma| - |\eta_\phi|)] \left(\frac{m_P}{\phi_\star}\right)^4 \left(\frac{r \mathcal{P}_\zeta}{8}\right)^2 \ln(kL) \,,
\end{eqnarray}
which reduces to
\begin{equation}
\left(\frac{\phi_\star}{m_P}\right)^4 = \left(\frac{r}{8}\right)^2 \mathcal{P}_\zeta \frac{\eta_\sigma^2}{\eta_\phi^4} \exp[4N(|\eta_\sigma| - |\eta_\phi|)] \ln(kL) \,, 
\end{equation}
where $\mathcal{P}_\zeta$ must be replaced by the observed value presented in Subsection \ref{observational}.


\subsubsection{$\zeta$ generated during inflation and $P_\zeta$ dominated by the tree-level term} \label{zipt}
According to Eq. (\ref{pt}), and the tensor to scalar ratio $r$ definition in Eq. (\ref{defr}), we have in this case
\begin{eqnarray}
\mathcal{P}_\zeta^{tree} &=& \frac{1}{\eta_\phi^2 \phi_\star^2} \left(\frac{H_\star}{2\pi}\right)^2 \nonumber \\
&=& \frac{1}{\eta_\phi^2} \left(\frac{m_P}{\phi_\star}\right)^2 \frac{r \mathcal{P}_\zeta}{8} \,, \label{cx1}
\end{eqnarray}
which reduces to
\begin{equation}
\left(\frac{\phi_\star}{m_P}\right)^2 = \frac{1}{\eta_\phi^2} \frac{r}{8} \,. \label{normt}
\end{equation}
Notice that in such a situation, the value of the $\phi$ field when the cosmologically relevant scales are exiting the horizon depends exclusively on the tensor to scalar ratio $r$, once $\eta_\phi$ has been fixed by the spectral tilt constraint as we will see later.


\subsection{Spectral tilt constraint} \label{sectilt}
The combined 5-year WMAP + Type I Supernovae + Baryon Acoustic Oscillations data \cite{komatsu1} constrain the value for the spectral tilt as
\begin{equation}
n_\zeta - 1 = -0.040 \pm 0.014 \,.
\end{equation}
Here again we have two possibilities:  $\mathcal{P}_\zeta$ is dominated either by the one-loop correction or by the tree-level term:

\subsubsection{$P_\zeta$ dominated by the one-loop correction} \label{indexkom}
In this case the usual spectral index formula at tree-level \cite{sasaki2} gets modified to account for the leading one-loop correction:
\begin{equation}
n_\zeta - 1 = -4\epsilon - 2m_P^2 \frac{\sum_{ijk} V_k N_{ijk} N_{ij}}{V \sum_{ij} N_{ij} N_{ij}} + \left[\ln (kL)\right]^{-1}\,.
\end{equation}
By making use of the derivatives in Eqs. (\ref{1d}), (\ref{2d}), and (\ref{3d}), we have
\begin{equation}
n_\zeta - 1 = -4\epsilon + 4\eta_\sigma + \left[\ln (kL)\right]^{-1}\,, \label{nskomat}
\end{equation}
which implies that the observed value for $n_\zeta$ is never reproduced 
in view of $\ln (kL) \sim \mathcal{O} (1)$.  Moreover, when calculating the running spectral index $dn_\zeta/d \ln k$ from Eq. (\ref{nskomat}), we obtain
\begin{equation}
\frac{dn_\zeta}{d \ln k} = -\left[\ln(kL)\right]^{-2} \,,
\end{equation}
which rules out the possibility that $P_\zeta$ is dominated by the one-loop correction since the calculated $dn_\zeta/d \ln k$ is far from the observationally allowed $2\sigma$ range of values:  $-0.0728 < dn_\zeta/d \ln k < 0.0087$ \cite{komatsu1}\footnote{We thank Eiichiro Komatsu for pointing out to us the dependence of $n_\zeta$ and $dn_\zeta/d\ln k$ on $\ln(kL)$.}.


\subsubsection{$P_\zeta$ dominated by the tree-level term}
Now the usual spectral index formula \cite{sasaki2} applies:
\begin{equation}
n_\zeta - 1 = -2\epsilon - 2m_P^2 \frac{\sum_{ij} V_i N_j N_{ij}}{V \sum_i N_i^2} \,,
\end{equation}
giving the following result once the derivatives in Eqs. (\ref{1d}), (\ref{2d}), and (\ref{3d}) have been used:
\begin{equation}
n_\zeta - 1 = -2\epsilon + 2\eta_\phi \,.
\end{equation}
The efect of the $\epsilon$ parameter may be discarded in the previous expression since, as often happens in small-field models \cite{alabidi2,boubekeur2}, $\epsilon$ is negligible being much less than $|\eta_\sigma|$:
\begin{equation}
\epsilon = \frac{m_P^2}{2} \frac{V_\phi^2 + V_\sigma^2}{V^2} = |\eta_\phi| \left[\frac{1}{2} |\eta_\phi| \left(\frac{\phi}{m_P}\right)^2\right] \ll |\eta_\phi| < |\eta_\sigma| \,,
\end{equation}
according to the prescription that the potential in Eq. (\ref{pot}) is dominated by the constant term.
Thus, by using the central value for $n_\zeta - 1$, we get
\begin{equation}
\eta_\phi = -0.020 \,. \label{tiltt}
\end{equation}

\subsection{Amount of inflation} \label{secamount}
It is well known that the number of e-folds of expansion from the time the cosmological scales exit the horizon to the end of inflation is presumably around but less than 62 \cite{liddle,dodelson}.  The slow-roll evolution of the $\phi$ field in Eq. (\ref{srp}) tells us that such an amount of inflation is given by
\begin{equation}
N = \frac{1}{|\eta_\phi|} \ln\left(\frac{\phi_{end}}{\phi_\star}\right) \lsim 62 \,. \label{end}
\end{equation}
Because of the characteristics of the inflationary potential in Eq. (\ref{pot}), there is no definite mechanism for ending inflation in this model.  It could not be by means of the violation of the $\epsilon < 1$ condition since this would imply extrapolating our results to a region where the potential in Eq. (\ref{pot}) is no longer dominated by the constant term which, in addition, would spoil the large non-gaussianity generated and could send the model to an unknowable quantum gravity regime.  Keeping in mind the results of Ref. \cite{armendariz2} which say that the ultraviolet cutoff in cosmological perturbation theory could be a few orders of magnitude bigger than $m_P$, we will therefore assume that inflation comes to an end when $|\eta_\phi|\phi^2/2m_P^2 \sim 10^{-2}$. This allows us to be on the safe side (avoiding large modifications to the potential coming from ultraviolet cutoff-suppressed non-renormalisable terms, and keeping the potential dominated by the constant $V_0$ term), leaving the implementation of a mechanism for ending inflation for a future work\footnote{We hope that the implementation of such a mechanism in our model will keep, or perhaps enhance, the generated non-gaussianity. Nevertheless the opposite behaviour might as well happen.  For instance, Ref. \cite{rigopoulos} studies within a stochastic formalism a quadratic two-component slow-roll model without a dominant constant term in the potential. A momentary violation of the slow-roll conditions around the end of inflation shows enhancement of $f_{NL}$ to observable levels; however, such an enhancement vanishes once inflation ends completely. These results have been confirmed numerically within the $\delta N$ formalism in Refs. \cite{vernizzi,yokoyama2}.}. Coming back to Eq. (\ref{end}), we get then
\begin{equation}
N = \frac{1}{|\eta_\phi|} \ln\left(\frac{m_P}{\phi_\star}\right) \lsim 62 \,, \label{amount}
\end{equation}
which leads to
\begin{equation}
\frac{\phi_\star}{m_P} \gsim \exp(-62|\eta_\phi|) \,. \label{amountc}
\end{equation}

\section{$f_{NL}$} \label{endcal}
In this section we will calculate the level of non-gaussianity represented in the parameter $f_{NL}$ by taking into account the constraints presented in Subsections \ref{secnorm}, \ref{sectilt}, and \ref{secamount}, and the different $\phi_\star$ regions discussed in Subsection \ref{secreg}.  


\subsection{The low $\phi_\star$ region}
This case is of no observational interest because $P_\zeta$ dominated by the one-loop correction is already ruled out by the observed spectral index and its running as shown in Subsubsection \ref{indexkom}.  In addition, the generated non-gaussianity is so big that it causes violation of the observational constraint $f_{NL} > -9$:
\begin{equation}
\frac{6}{5} f_{NL} = \frac{B_\zeta^{1-loop}}{4\pi^4 \frac{\sum_i k_i^3}{\prod_i k_i^3} (\mathcal{P}_\zeta^{1-loop})^2} = -[\mathcal{P}_\zeta \ln(kL)]^{-1/2} \sim -2 \times 10^4 \,,
\end{equation}
according to the expressions in Eqs. (\ref{asidef}), (\ref{bfp}), (\ref{pl}), and (\ref{bl}). 

We want to remark that, although it is of no observational interest, this case represents the first example of large non-gaussianity in the bispectrum $B_\zeta$ of $\zeta$ for a slow-roll model of inflation with canonical kinetic terms. It is funny to realize that the model in this case is additionally ruled out because the observational constraint on $f_{NL}$ is violated {\it by an excess} and not by a shortfall as is currently thought \cite{vernizzi,battefeld,yokoyama1,maldacena,seery7}.

\subsection{The intermediate $\phi_\star$ region}

The level of non-gaussianity, according to the expressions in Eqs. (\ref{asidef}), (\ref{bfp}), (\ref{pt}), and (\ref{bl}), is in this case given by
\begin{eqnarray}
\frac{6}{5} f_{NL} &=& \frac{B_\zeta^{1-loop}}{4\pi^4 \frac{\sum_i k_i^3}{\prod_i k_i^3} (\mathcal{P}_\zeta^{tree})^2} = \frac{\eta_\sigma^3}{\eta_\phi^2 \phi_\star^2} \exp[6N(|\eta_\sigma|-|\eta_\phi|)] \left(\frac{H_\star}{2\pi}\right)^2 \ln(kL) \nonumber \\
&=& \frac{\eta_\sigma^3}{\eta_\phi^2} \exp[6N(|\eta_\sigma|-|\eta_\phi|)] \left(\frac{m_P}{\phi_\star}\right)^2 \frac{r \mathcal{P}_\zeta}{8} \ln(kL) \nonumber \\
&=& \eta_\sigma^3 \exp[6N(|\eta_\sigma|-|\eta_\phi|)] \mathcal{P}_\zeta \ln(kL) \,, \label{poseta} \\
\Rightarrow \frac{6}{5} f_{NL} &\approx& -2.457 \times 10^{-9} |\eta_\sigma|^3 \exp[300 \ \ln(5.657 \times 10^{-2} r^{-1/2}) \left(|\eta_\sigma| - 0.020\right)] \,,
\end{eqnarray}
where in the last line we have used expressions from Eqs. (\ref{normt}), (\ref{tiltt}), and (\ref{amount}).

Now, by implementing the spectral tilt constraint in Eq. (\ref{tiltt}) in the spectrum normalisation constraint in Eq. (\ref{normt}) and the amount of inflation constraint in Eq. (\ref{amountc}), we conclude that the tensor to scalar ratio $r$ is bounded from below:  $r \gsim 2.680 \times 10^{-4}$. 

\begin{figure*} [t]
\begin{center}
\includegraphics[width=15cm,height=10cm]{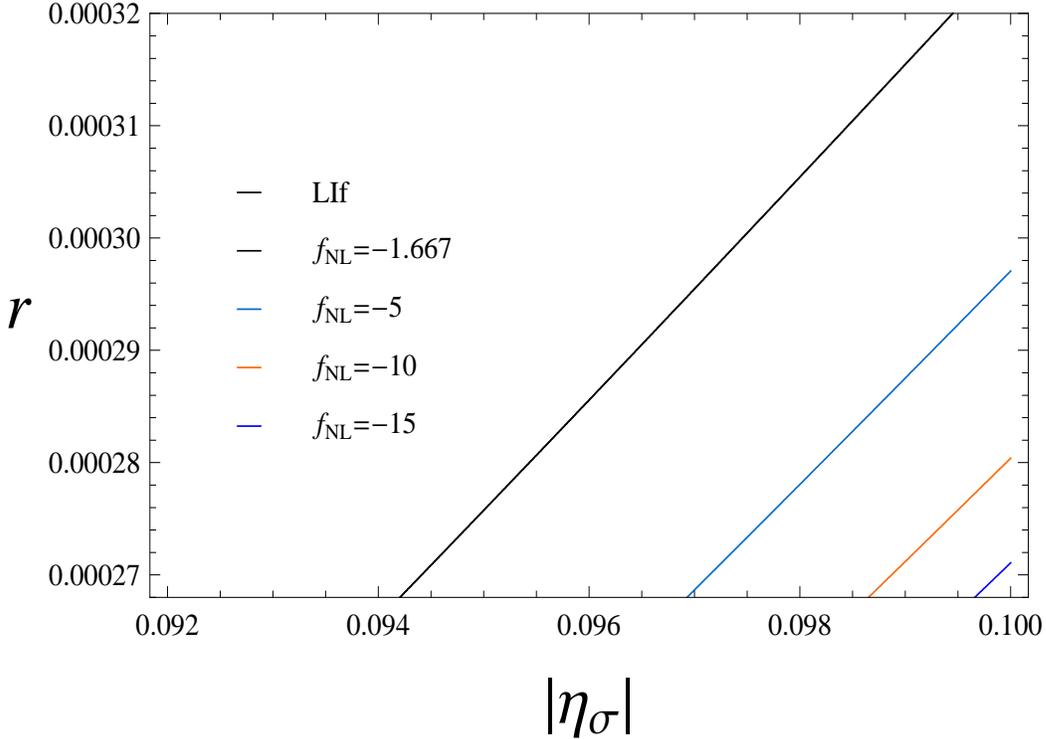}
\end{center}
\caption{Contours of $f_{NL}$ in the $r$ vs $|\eta_\sigma|$ plot.  
The intermediate (high) $\phi_\star$ region 
is below (above) the LIf line. 
The WMAP (and also PLANCK) observationally allowed $2\sigma$ range of values for negative $f_{NL}$, $-9 < f_{NL}$, is completely inside the intermediate $\phi_\star$ region.  Notice that the LIf line matches almost exactly the $f_{NL} = -1.667$ line.}
\label{fig1}
\end{figure*}

In the plot $r$ vs $|\eta_\sigma|$ in figure \ref{fig1}, we show lines of constant $f_{NL}$ corresponding to the values $f_{NL} = -5, -10, -15$.  We also show the high and intermediate $\phi_\star$ regions in agreement with the constraint in Eq. (\ref{intc}):
\begin{eqnarray}
&&\frac{r \mathcal{P}_\zeta}{8} \frac{\eta_\sigma^2}{\eta_\phi^2} \exp[4N(|\eta_\sigma|-|\eta_\phi|)] \ll \left(\frac{\phi_\star}{m_P}\right)^2 \ll \frac{r \mathcal{P}_\zeta}{8} \frac{\eta_\sigma^3}{\eta_\phi^3} \exp[6N(|\eta_\sigma|-|\eta_\phi|)] \,, \nonumber \\
&\Rightarrow& 8.139 \times 10^6 \ll |\eta_\sigma|^3 \exp[300 \ln(5.657 \times 10^{-2} r^{-1/2})\left(|\eta_\sigma| - 0.020\right)] \ll 8.210 \times 10^{12} \,.
\end{eqnarray}
As is evident from the plot, the WMAP (and also PLANCK) observationally allowed $2\sigma$ range of values for negative $f_{NL}$, $-9 < f_{NL}$, is completely inside the intermediate $\phi_\star$ region as required.  More negative values for $f_{NL}$, up to $f_{NL} = -20.647$ are consistent within our framework for the intermediate $\phi_\star$ region, but they are ruled out from observation. Nevertheless, like for the low $\phi_\star$ region studied above, it is interesting to see a slow-roll inflationary model with canonical kinetic terms where the observational restriction on $f_{NL}$ may be violated {\it by an excess} and not by a shortfall. So we conclude that {\it if} $B_\zeta$ {\it is dominated by the one-loop correction but} $P_\zeta$ {\it is dominated by the tree-level term, sizeable non-gaussianity is generated even if} $\zeta$ {\it is generated during inflation}.  We also conclude, from looking at the small values that the tensor to scalar ratio $r$ takes in figure \ref{fig1} compared with the present technological bound $r \gsim 10^{-3}$ \cite{friedman}, that {\it for non-gaussianity to be observable in this model, primordial gravitational waves must be undetectable}.

Notice that in oder to get positive values for $f_{NL}$, which is observationally more interesting in view of the results presented in Subsection \ref{observational}, $\eta_\sigma$ should be positive according to Eq. (\ref{poseta}).  However, being $\eta_\phi$ negative in order to reproduce the observed spectral tilt, the argument of the exponential in Eq. (\ref{poseta}) would be negative, making the $f_{NL}$ obtained too small to be observationally interesting\footnote{We thank Eiichiro Komatsu for questioning us about this issue.}.  As regards the general case, in view of the previous reason being model dependent, we may only say that in order to get $f_{NL}$ positive when $B_\zeta$ is dominated by the one-loop corrections, $B_\zeta$ should be positive (based on the definition of $f_{NL}$ in Eq. (\ref{bfp})) which means that the maximum between $N_{\sigma \sigma}$ and $N_{\phi \phi}$ should be positive in view of Eq. (\ref{poskomb}).

Finally we want to point out that, by reducing our model to the single-field case, the consistency relation between $f_{NL}$ and $n_\zeta$ presented in Ref. \cite{creminelli}: $f_{NL} \sim \mathcal{O} (n_\zeta -1)$ is not violated since in that case $B_\zeta$ is never dominated by the one-loop corrections for slow-roll inflation as demonstrated in Ref. \cite{lyth1}\footnote{We thank Filippo Vernizzi for questioning us about this issue.}. Thus, the level of non-gaussianity $f_{NL}$ for our model reduced to the single-field case is described by the high $\phi_\star$ region as shown below.

\subsection{The high $\phi_\star$ region}
This case is of no observational interest because, according to the expressions in Eqs. (\ref{asidef}), (\ref{bfp}), (\ref{pt}), (\ref{bt}), and (\ref{tiltt}), the non-gaussianity generated is too small to be observable:
\begin{equation}
\frac{6}{5} f_{NL} = \frac{B_\zeta^{tree}}{4\pi^4 \frac{\sum_i k_i^3}{\prod_i k_i^3} (\mathcal{P}_\zeta^{tree})^2} = -\eta_\phi = 0.020 \,,
\end{equation}
in agreement with the consistency relation of Ref. \cite{creminelli} for our model reduced to the single-field case, and with the general expectations of Refs. \cite{vernizzi,battefeld,yokoyama1,maldacena,seery7} for slow-roll inflationary models with canonical kinetic terms where only the tree-level contributions are considered.

\section{Convergence of the $\zeta$ series and perturbative regime} \label{seccou}
In Sections \ref{rest} and \ref{endcal} we have worked up to the one-loop diagrams in order to constrain the parameter space and find the level of non-gaussianity $f_{NL}$. It is time then to address the issue of the $\zeta$ series convergence and justify the existence of a perturbative regime so that the truncation of the series up to the one-loop order, for the model we have considered, is valid. A way to do that is by rederiving the $\zeta$ series in terms of $\delta \phi_\star$ and $\delta \sigma_\star$ by equating the unperturbed scalar potential to the perturbed one at the final time $t$;  this of course is valid in view of the first slow-roll condition in Eq. (\ref{1stsrc}) and the final slice being one of uniform energy density:
\begin{eqnarray}
&&V_0 \left \{1 + \frac{1}{2} \eta_\phi \frac{\phi_\star^2}{m_P^2} \exp[-2N\eta_\phi] + \frac{1}{2} \eta_\sigma \frac{\sigma_\star^2}{m_P^2} \exp[-2N\eta_\sigma] \right \} \nonumber \\
&&= V_0 \left \{1 + \frac{1}{2} \eta_\phi \frac{(\phi_\star + \delta \phi_\star)^2}{m_P^2} \exp[-2(N+\delta N)\eta_\phi] + \frac{1}{2} \eta_\sigma \frac{(\sigma_\star + \delta \sigma_\star)^2}{m_P^2} \exp[-2(N + \delta N)\eta_\sigma] \right \} \,. \nonumber \\
&&
\end{eqnarray}
From the previous expression it follows that
\begin{eqnarray}
&&\eta_\phi \phi_\star^2 \exp[-2N\eta_\phi] + \eta_\sigma \sigma_\star^2 \exp[-2N\eta_\sigma] \nonumber \\
&&= \eta_\phi (\phi_\star + \delta \phi_\star)^2 \exp[-2(N+\delta N)\eta_\phi] + \eta_\sigma (\sigma_\star + \delta \sigma_\star)^2 \exp[-2(N + \delta N)\eta_\sigma] \,,  \label{1con}
\end{eqnarray}
which is easier to handle in terms of variables $x$ and $y$ defined as
\begin{eqnarray}
x &\equiv& \frac{\delta \phi_\star}{\phi_\star} \,, \label{xvdef} \\
y &\equiv& \left[\frac{\eta_\sigma^3}{\eta_\phi^3} \frac{\sigma_\star^2}{\phi_\star^2} \left(1 + \frac{\delta \sigma_\star}{\sigma_\star}\right)^2 \exp[2N(\eta_\phi-\eta_\sigma)]\right]^{1/2} \,. \label{yvdef}
\end{eqnarray}
Thus, the exponentials factors contaning $N$ (but not $\delta N$) are completely absorbed in $y$, and the expression in Eq. (\ref{1con}) looks as follows:
\begin{equation}
1 + \frac{\eta_\phi^2}{\eta_\sigma^2} \frac{1}{\left(1 + \frac{\delta \sigma_\star}{\sigma_\star}\right)^2} y^2 = (1 + x)^2 \exp [-2 \delta N \eta_\phi] + \frac{\eta_\phi^2}{\eta_\sigma^2} y^2 \exp[-2 \delta N \eta_\sigma] \,. \label{xyme}
\end{equation}

If we were able to solve for $\delta N$ in Eq. (\ref{xyme}) in terms of $\eta_\phi$, $\eta_\sigma$, $x$, and $y$ (after making $\sigma_\star = 0$), we could Taylor-expand around $x=0$ and $y=0$ reproducing the ${\bf x}$-dependent part of Eq. (\ref{Nexp}).  This would be really good because the Taylor expansion would look so clean, in the sense that all the concerning exponential factors contaning $N$ which appear explicitely in Eq. (\ref{Nexp}) would already be absorbed in $y$, that the issue of truncating at some specific order in $\delta \phi_\star$ and $\delta \sigma_\star$ would be simply justified by requiring $|x| \ll 1$ and $|y| \ll 1$.
Nevertheless, as is seen in Eq. (\ref{xyme}), it is impossible to solve for $\delta N$ in terms of $\eta_\phi$, $\eta_\sigma$, $x$, and $y$ unless we make a Taylor expansion of the exponential functions aroud $\delta N = 0$: 
\begin{eqnarray}
0 &=&  \left \{\left[(1 + x)^2 - 1\right] + \frac{\eta_\phi^2}{\eta_\sigma^2} y^2 \left[1 - \frac{1}{\left(1 + \frac{\delta \sigma_\star}{\sigma_\star} \right)^2} \right] \right \}+ \nonumber \\
&&+\delta N  \left[-2\eta_\phi (1 + x)^2 - 2\frac{\eta_\phi^2}{\eta_\sigma} y^2 \right] + \delta N^2  \left[2\eta_\phi^2 (1 + x)^2 + 2\eta_\phi^2 y^2 \right] + ... \,. \label{2con}
\end{eqnarray}
Notice that the Taylor expansion of the exponential functions is always convergent whatever the arguments of the exponentials are.  Moreover, if the Taylor expansion derived from a function $f(x)$ converges, it converges precisely to $f(x)$ \cite{spivak}.  Thus, the expression in Eq. (\ref{2con}) is actually the same as the expression in Eq. (\ref{xyme}).

Now, solving for $\delta N$ in terms of $\eta_\phi$, $\eta_\sigma$, $x$, and $y$, although possible in view of the expression in Eq. (\ref{2con}), is not an easy business. That is why we will truncate the series in Eq. (\ref{2con}) up to second order in $\delta N$ and solve the resultant quadratic equation\footnote{The truncation up to second order in $\delta N$ has been chosen in order to have complete consistency with the order of the variables $x$ and $y$ in Eq. (\ref{xyme}).}. Notice that, since $\zeta \equiv \delta N - \langle \delta N \rangle$ and $\zeta \sim 10^{-5}$, we may truncate the series in Eq. (\ref{2con}) up to whatever order we wish and still reproduce $\zeta$ to high accuracy.  
Thus, the solution for the quadratic equation coming from the series in Eq. (\ref{2con}) after truncation at second order is:
\begin{eqnarray}
\delta N &\approx& \Big \{ \frac{1}{2} \left(1 + x \right)^2 + \frac{1}{2} \frac{\eta_\phi}{\eta_\sigma} y^2 \pm \Big \{ \left[\frac{1}{2} \left(1 + x \right)^2 + \frac{1}{2} \frac{\eta_\phi}{\eta_\sigma} y^2 \right]^2 - \nonumber \\
&& - \frac{1}{2} \left \{ \left[\left(1 + x \right)^2 - 1\right] + \frac{\eta_\phi^2}{\eta_\sigma^2} y^2 \left[1 - \frac{1}{\left(1 + \frac{\delta \sigma_\star}{\sigma_\star}\right)^2} \right] \right \} \left[\left(1 + x \right)^2 + y^2  \right] \Big \}^{1/2} \Big \} \times \nonumber \\
&& \times \left \{\eta_\phi \left[ \left(1 + x \right)^2 + y^2 \right] \right \}^{-1} \,. \label{dnxy}
\end{eqnarray}
If in addition we make Taylor expansions of the square root and the factor in the third line of the previous expression around $x=0$ and $y=0$:
\begin{eqnarray}
&&\left \{ \left[\frac{1}{2} \left(1 + x \right)^2 + \frac{1}{2} \frac{\eta_\phi}{\eta_\sigma} y^2 \right]^2 - \frac{1}{2} \left \{ \left[\left(1 + x \right)^2 - 1\right] + \frac{\eta_\phi^2}{\eta_\sigma^2} y^2 \left[1 - \frac{1}{\left(1 + \frac{\delta \sigma_\star}{\sigma_\star}\right)^2} \right] \right \} \left[\left(1 + x \right)^2 + y^2  \right] \right \}^{1/2} \nonumber \\
&&= \frac{1}{2} - x^2 + \frac{\eta_\phi}{2\eta_\sigma} \left \{ 1 - \frac{\eta_\phi}{\eta_\sigma} \left[1 + \frac{1}{\left(1 + \frac{\delta \sigma_\star}{\sigma_\star}\right)^2} \right] \right \} y^2 + ... \,, \label{sqrtexp} \\
&&\left \{\eta_\phi \left[ \left(1 + x \right)^2 + y^2 \right] \right \}^{-1} = \frac{1}{\eta_\phi} \left[1 - 2x + 3x^2 -y^2 + ... \right] \,, \label{-1exp}
\end{eqnarray}
introducing them into Eq. (\ref{dnxy}), we end up with the following power series for $\delta N$:
\begin{equation}
\delta N \approx \frac{1}{\eta_\phi} \left( x - \frac{x^2}{2} + \frac{\eta_\phi^2}{2\eta_\sigma^2} y^2 + ... \right) \,,  \label{dnexxy}
\end{equation}
where the $\pm$ symbol is changed to the $-$ sign so that $\delta N$ remains a perturbation, and the trajectory $\sigma = 0$ is chosen. Coming back to the variables $\delta \phi_\star$ and $\delta \sigma_\star$ we see that Eq. (\ref{dnexxy}) reproduces the ${\bf x}$-dependent part of Eq. (\ref{Nexp}) in view of Eqs. (\ref{1d}) and (\ref{2d}) up to second order in $\delta \phi_\star$ and $\delta \sigma_\star$:
\begin{equation}
\delta N \approx \frac{\delta \phi_\star}{\eta_\phi \phi_\star} - \frac{1}{2\eta_\phi} \left(\frac{\delta \phi_\star}{\phi_\star}\right)^2 + \frac{\eta_\sigma}{2\eta_\phi^2} \left(\frac{\delta \sigma_\star}{\phi_\star}\right)^2 \exp \left[2N \left(\eta_\phi - \eta_\sigma \right) \right] + ... \,. 
\end{equation}

Eq. (\ref{dnexxy}), although reliable only up to second order, tells us that the expected behaviour of $\delta N$ in terms of $\eta_\phi$, $\eta_\sigma$, $x$, and $y$ is indeed obtained.  Moreover, from our previous discussion we know that $\delta N$ can be exactly written in terms of a series of $x$ and $y$ withouth the explicit appearance of the concerning exponential factors containing $N$. This is indeed partially confirmed up to third order when introducing Eqs. (\ref{1d}), (\ref{2d}), and (\ref{3d}) into the ${\bf x}$-dependent part of Eq. (\ref{Nexp}):
\begin{equation}
\delta N =\frac{1}{\eta_\phi} \left( x - \frac{x^2}{2} + \frac{\eta_\phi^2}{2\eta_\sigma^2} y^2 + \frac{x^3}{3} - \frac{\eta_\phi}{3\eta_\sigma} xy^2 + ... \right) \,.  \label{dntrun}
\end{equation}

The bottom line of this discussion is that we have been able to identify two quantities that determine the truncation of the series up to some specific order.  These two quantities are $x$ and $y$ which we could identify as the ``coupling constants'' of the theory in the context of Quantum Field Theory.  By making $|x| \ll 1$ and $|y| \ll 1$ we can see from Eq. (\ref{dntrun}) that all the terms higher than second order in $x$ and $y$ are subleading compared to the second-order ones.  As regards the first-order terms compared to the second-order ones, we see that the latter are not necessarily subleading compared to the former because of the non-existence of the first-order $y$ term and in view of $|y/x| \lsim 1600$ from Eqs. (\ref{xvdef}) and (\ref{yvdef}) and the values for $\eta_\phi$, $\eta_\sigma$ and $N$ considered in Sections \ref{rest} and \ref{endcal}. In the language of the Feynman-like diagrams \cite{byrnes1}, truncating $\delta N$ in Eq. (\ref{dntrun}) up to second order in $x$ and $y$ means considering only the leading diagrams at tree level and one loop which is what we have done in Sections \ref{rest} and \ref{endcal}. In fact, $|x| \ll 1$ and $|y| \ll 1$ mean that
\begin{eqnarray}
|x| &\equiv& \left|\frac{\delta \phi_\star}{\phi_\star}\right| \approx \left(\frac{H_\star}{2\pi} \right) \frac{1}{\phi_\star} \ll 1 \,,  \\
|y| &\equiv& \left \{ \frac{\eta_\sigma^3}{\eta_\phi^3} \frac{\delta \sigma_\star^2}{\phi_\star^2} \exp [2N (\eta_\phi - \eta_\sigma)] \right \}^{1/2} \approx \left \{ \frac{\eta_\sigma^3}{\eta_\phi^3} \left(\frac{H_\star}{2\pi}\right)^2 \frac{1}{\phi_\star^2} \exp [2N (\eta_\phi - \eta_\sigma)] \right \}^{1/2} \ll 1 \,,
\end{eqnarray}
which are well satisfied for the cases when $P_\zeta$ is dominated by the tree-level term (see Subsubsection \ref{zipt} - Eq. (\ref{cx1}) and Subsubsection \ref{bz1l} - Eq. (\ref{wn})):
\begin{eqnarray}
&&\left(\frac{H_\star}{2\pi} \right) \frac{1}{\phi_\star} = |\eta_\phi| \mathcal{P}_\zeta^{1/2} \approx 10^{-6} \,, \\
&&\left \{ \frac{\eta_\sigma^3}{\eta_\phi^3} \left(\frac{H_\star}{2\pi}\right)^2 \frac{1}{\phi_\star^2} \exp [2N (\eta_\phi - \eta_\sigma)] \right \}^{1/2} \ll \left \{ \frac{\eta_\sigma}{\eta_\phi} \exp [-2N (|\eta_\sigma| - |\eta_\phi|)] \right \}^{1/2}  \lsim 2 \,.
\end{eqnarray}
By explicitly calculating the two-loop and three-loop diagrams for $P_\zeta$ and $B_\zeta$, and employing the results of Ref. \cite{jarnhus}, we have checked that the conditions $|x| \ll 1$ and $|y| \ll 1$ efectively make these diagrams subleading compared to the leading ones at one-loop level.

Finally, we will discuss the convergence of the $\zeta$ series in view of Eqs. (\ref{dnxy}), (\ref{sqrtexp}), and (\ref{-1exp}).  We first note that the series in Eq. (\ref{sqrtexp}) is always convergent.  As regards the series in Eq. (\ref{-1exp}), it will not be convergent at all while the function
\begin{equation}
\frac{1}{(1+x)^2 + y^2} \approx \frac{1}{(1+x)^2 + B^2 x^2} \,,  \label{fB}
\end{equation}
with
\begin{equation}
B = \left(\frac{\eta_\sigma}{\eta_\phi}\right)^{3/2} \exp [N(\eta_\phi - \eta_\sigma)] \,, 
\end{equation}
does not satisfy the following necessary condition \cite{spivak}:  for the Taylor series around $x=0$ of a function $f(x)$ to be convergent, it is necessary that the extension $f(z)$ to the complex plane of $f(x)$ is continous in a neighbourhood of $z=0$.  If this is the case, and the Taylor series of $f(z)$ is indeed convergent, the convergence circle must either match or be inside the aforementioned neighbourhood. Of course, this is not a sufficient condition, but at least gives us a constraint on the possible values that $x$ may take.

Applying this condition to the expression in Eq. (\ref{fB}), we see that the extension of this function to the complex plane has poles for $(1+z)^2 = -B^2 z^2$ which leads to
\begin{equation}
z = \frac{\pm iB - 1}{B^2 + 1} \,. 
\end{equation}
Therefore, the extension to the complex plane of Eq. (\ref{fB}) is continous for
\begin{equation}
|z| < \frac{B^2 + 1}{B^2 + 1} = 1\,, 
\end{equation}
so the necessary condition for the convergence of the series in Eq. (\ref{-1exp}), and therefore for the convergence of the series in Eq. (\ref{dnxy}) which is what we are interested in, is given by $|x| < 1$.  Thus, such a necessary condition for the convergence of the $\zeta$ series is automatically satisfied once we choose $|x| \ll 1$, as we have seen above it is required for working in a perturbative regime.

\section{Conclusions} \label{concl}
Observational cosmology is in its golden age: current satellite and balloon experiments are working extremely well \cite{wmap,hinshaw}, dramatically improving the quality of data \cite{komatsu1}.  Moreover, foreseen experiments \cite{planck,planck1} will take the field to a state of unprecedent precission where theoretical models will be subjected to the most demanding tests. Given such a state of affairs, it is essential to study the higher order statistical descriptors for cosmological quantities such as the primordial curvature perturbation $\zeta$, which give us information about the non-gaussianity in their corresponding probability distribution functions.

$\zeta$ and its associated non-gaussianity depend on the specific inflationary model that describes the dynamics of the early Universe, the slow-roll class of inflationary models with canonical kinetic terms being the most popular and studied to date. Inflationary models of the slow-roll variety predict very well the spectral index in the spectrum $P_\zeta$ of $\zeta$ but, if the kinetic terms are canonical, they seem to generate unobservable levels of non-gaussianity in the bispectrum $B_\zeta$ and the trispectrum $T_\zeta$ of $\zeta$ making them impossible to test against the astonishing forthcoming data. Where does this conclusion come from? The answer relies on careful calculations of the levels of non-gaussianity $f_{NL}$ and $\tau_{NL}$ by making use of the $\delta N$ formalism \cite{vernizzi,battefeld,yokoyama1,seery3}. In this framework, $\zeta$ is given in terms of the perturbation $\delta N$ in the amount of expansion from the time the cosmologically relevant scales exit the horizon until the time at which one wishes to calculate $\zeta$.

Due to the functional dependence of the amount of expansion, $\zeta$ is usually Taylor-expanded (see Eq. (\ref{Nexp})) and truncated up to some desired order so that $f_{NL}$ and $\tau_{NL}$ are easily calculated (see for instance Eq. (\ref{fdnf})). Two key questions arise when noting that it is impossible to extract general and useful information from the $\zeta$ series expansion in Eq. (\ref{Nexp}) until one chooses a definite inflationary model and calculates explicitly the $N$ derivatives. First of all, when writing a general expression for $f_{NL}$ or $\tau_{NL}$ in terms of the $N$ derivatives, how do we know that such an expression is correct if the series convergence has not been examined? Moreover, if the convergence radius of the $\zeta$ series is already known, why is each term is the $\zeta$ series supposed to be smaller than the previous one so that cutting the series at any desired order is thought to be enough to keep the leading terms? Nobody seems to have formulated these questions before and, by following a naive line of thinking, $f_{NL}$ and $\tau_{NL}$ were calculated for slow-roll inflationary models with canonical kinetic terms without checking the $\zeta$ series convergence and keeping only the presumably leading tree-level terms \cite{vernizzi,battefeld,yokoyama1,seery3,maldacena,seery7}.

These two questions have been addressed in this paper by paying attention to a particular quadratic small-field slow-roll model of inflation with two components and canonical kinetic terms (see Eq. (\ref{pot})). Although the non-diagrammatic approach followed in Section \ref{seccou} to find the necessary condition for the convergence of the $\zeta$ series in our model might not be applicable to all the cases, we have been able to show that not being careful enough when choosing the right available parameter space could make the $\zeta$ series, and therefore the calculation of $f_{NL}$ and $\tau_{NL}$ from the truncated series (e.g. Eq. (\ref{fdnf})), meaningless. We also have been able to show in our model that the one-loop terms in the spectrum $P_\zeta$ and/or the bispectrum $B_\zeta$ of $\zeta$ could be bigger or lower than the corresponding tree-level terms, but are always much bigger than the corresponding terms whose order is higher than the one-loop order. If both $P_\zeta$ and $B_\zeta$ are dominated by the one-loop terms, {\it a huge} $f_{NL}$ {\it is generated} which overwhelms the observational constraint, ruling out the model {\it by an excess} and not by a shortfall. If $B_\zeta$ is still dominated by the one-loop correction but $P_\zeta$ is now dominated by the tree-level term, {\it sizeable and observable values for} $f_{NL}$ {\it are generated}, so they can be tested against present and forthcoming observational data. Finally, if both $P_\zeta$ and $B_\zeta$ are dominated by the tree-level terms, $f_{NL}$ {\it is slow-roll suppressed} as was originally predicted in Refs. \cite{vernizzi,battefeld,yokoyama1}.

What these results teach us is that the issue of the $\zeta$ series convergence and loop corrections is essential for making correct predictions about the statistical descriptors of $\zeta$ in the framework of the $\delta N$ formalism, and promising for finding high levels of non-gaussianity that can be compared with observations. In fact, now that we have learned the lesson, the level of non-gaussianity $\tau_{NL}$ for the same slow-roll model studied here will be the subject of a companion paper \cite{cogollo}.

\bigskip

\subsection*{Acknowledgments} 
This work was supported by COLCIENCIAS grant No. 1102-333-18674 CT-174-2006 and by DIEF (UIS) grant No. 5134. H.R.S.C. acknowledges Fundaci\'on Mazda para el Arte y la Ciencia for a postgraduate scholarship. Y.R. thanks Bartjan van Tent, David H. Lyth, Antonio Riotto, Eiichiro Komatsu, Filippo Vernizzi, Christian Byrnes, Lotfi Boubekeur, and Juli\'an Jaimes for useful comments and stimulating discussions. Y.R. also acknowledges the Laboratoire de Physique Th\'eorique d'Orsay at the Universit\'e de Paris Sud (France) where part of this work was done with the support of the ECOS Programme grant No. C06P02, as well as the Department of Physics at Lancaster University (UK) for hospitality during the time he spent there working on this subject. Special thanks to MSO(2007) for being the most wonderful and indeed unique source of motivation and inspiration in all the stages of this work; this paper is dedicated to her memory.

\appendix 

\section{Tree-level and one-loop diagrams for $P_\zeta$ and $B_\zeta$} \label{app}

We show in this appendix the mathematical expressions for the tree-level and one-loop Feynman-like diagrams associated with the spectrum $P_\zeta$ and the bispectrum $B_\zeta$ of $\zeta$, following the set of rules presented in Ref. \cite{byrnes1}. To this end we have taken into account the $N$ derivatives for our small-field slow-roll model given in Eqs. (\ref{1d}), (\ref{2d}), and (\ref{3d}).  After presenting the mathematical expressions, we will estimate the order of magnitude of each diagram in order to determine the respective leading terms at tree-level and one-loop for both $P_\zeta$ and $B_\zeta$.

\subsection{Tree-level diagram for $P_\zeta$} \label{diagramsP}

\begin{figure}
\begin{center}
\includegraphics[width=7cm]{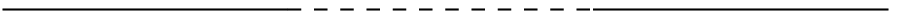}
\end{center}
\caption{Tree-level Feynman-like diagram for $P_\zeta$. The internal dashed line corresponds to a two-point correlator of field perturbations.} \label{tlpd}
\end{figure}

Looking at Fig. \ref{tlpd}, we see that $P_\zeta^{tree}$ is given by
\begin{eqnarray}
P_\zeta^{tree} &=& N_\phi^2 \ P_{\delta \phi} (k) \nonumber \\
&=& \frac{2\pi^2}{k^3} \frac{1}{\eta_\phi^2 \phi_\star^2} \left(\frac{H_\star}{2\pi}\right)^2 \,. \label{pt1}
\end{eqnarray}
Of course, there is only one tree-level diagram for $P_\zeta$ and therefore Eq. (\ref{pt1}) is the associated leading tree-level term.

Our calculation in this appendix goes up to the one-loop diagrams so, in order to have complete consistency in the calculation \cite{seery2}, we should also take into account the one-loop correction to the two-point correlator in the field perturbations when calculating the diagram in Fig. \ref{tlpd}. Such a correction has been studied in Refs. \cite{weinberg1,weinberg2,sloth1,sloth2,seery1} where the most general result for single-field slow-roll inflation with $N_{total}$ not very much bigger than 62 is \cite{seery1}
\begin{equation}
P_{\delta \phi}^{1-loop} = \frac{2\pi^2}{k^3} \left( \frac{H_\star}{2\pi}\right)^2 \left \{1 + \left(\frac{H_\star}{2\pi m_P}\right)^2 \left[\frac{35}{6} \ln(kL) + \beta \right] \right \} \,, \label{q1l}
\end{equation}
where $L$ is the infrared cutoff for a minimal box \cite{lyth1,bernardeu4}, and $\beta$ is a renormalisation scheme-dependent constant that is expected to be negligible on large scales compared to $\ln (kL) \sim \mathcal{O}(1)$.  The one-loop correction to the field perturbation spectrum in Eq. (\ref{q1l}) is, therefore, negligible compared to the tree-level contribution $P_{\delta \phi}^{tree} = (2\pi^2/k^3) (H_\star/2\pi)^2$ if $H_\star \ll m_P$ as usually required.  In our model $H_\star \ll m_P$ is indeed given but, since we are dealing with a two-component model, the result in Eq. (\ref{q1l}) may not be applicable. Anyway, we feel quite confident that the (up to now unknown) extension of Eq. (\ref{q1l}) to the multiple-field case will yield similar results, so we will keep the expression in Eq. (\ref{pt1}) as the leading tree-level contribution to $P_\zeta$.

\subsection{One-loop diagrams for $P_\zeta$} \label{Ldis}

\begin{figure}
\begin{center}
\begin{tabular}{cccc}
\includegraphics[width=7cm,height=1cm]{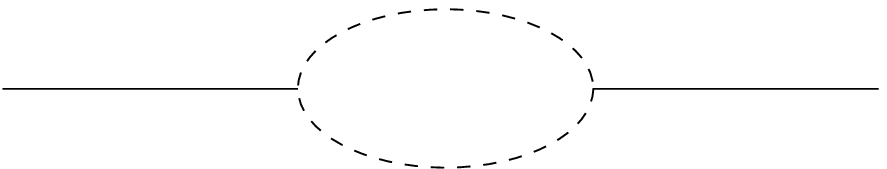} & & & \includegraphics[width=7cm,height=1cm]{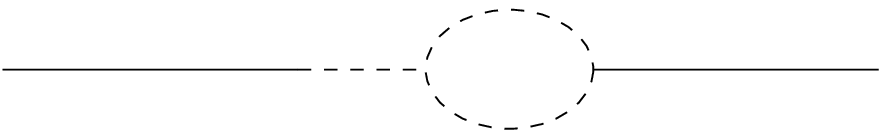} \\
(a) & & & (b)
\end{tabular}
\end{center}
\caption{One-loop Feynman-like diagrams for $P_\zeta$. (a). The two internal dashed lines correspond to two-point correlators of field perturbations. (b). The internal dashed lines correspond to a three-point correlator of field perturbations.} \label{olpd}
\end{figure}

Looking at Figs. \ref{olpd}a and \ref{olpd}b, we see that $P_\zeta^{1-loop}$ is given by two contributions $P_\zeta^{1-loop \; a}$ and $P_\zeta^{1-loop \; b}$:
\begin{eqnarray}
P_\zeta^{1-loop \; a} &=& \frac{1}{2} \left[N_{\phi \phi}^2 + N_{\sigma \sigma}^2\right] \int \frac{d^3 q}{(2\pi)^3} P_{\delta \phi}(q) P_{\delta \phi} (|{\bf k} + {\bf q}|) \nonumber \\
&=& \frac{1}{2} \left[\frac{1}{\eta_\phi^2 \phi_\star^4} + \frac{\eta_\sigma^2}{\eta_\phi^4 \phi_\star^4} \exp\left[4N(\eta_\phi - \eta_\sigma)\right]\right] \frac{4\pi^2}{k^3} \ln(kL) \left(\frac{H_\star}{2\pi}\right)^4 \,, \label{1la} \\
P_\zeta^{1-loop \; b} &=& N_\phi N_{\phi \phi} \int \frac{d^3 q}{(2\pi)^3} B_{\delta \phi \ \delta \phi \ \delta \phi} (k,q,|{\bf k} + {\bf q}|) + \nonumber \\
&& + N_\phi N_{\sigma \sigma} \int \frac{d^3 q}{(2\pi)^3} B_{\delta \phi \ \delta \sigma \ \delta \sigma} (k,q,|{\bf k} + {\bf q}|) \nonumber \\
&=& -\frac{1}{\eta_\phi^2 \phi_\star^3} \left[\int \frac{d^3 q}{(2\pi)^3} 4\pi^4 \sum_{perm} \left(\frac{H_\star}{2\pi}\right)^4 \frac{\epsilon_\phi^{1/2}}{2\sqrt{2} m_P} \frac{\mathcal{M}_3(k,q,|{\bf k} + {\bf q}|)}{k^3 q^3 |{\bf k} + {\bf q}|^3} \right] + \nonumber \\
&& + \frac{\eta_\sigma}{\eta_\phi^3 \phi_\star^3} \exp\left[2N(\eta_\phi - \eta_\sigma)\right] \left[\int \frac{d^3 q}{(2\pi)^3} 4\pi^4 \sum_{perm. \ l2a.} \left(\frac{H_\star}{2\pi}\right)^4 \frac{\epsilon_\phi^{1/2}}{2\sqrt{2} m_P} \frac{\mathcal{M}_3(k,q,|{\bf k} + {\bf q}|)}{k^3 q^3 |{\bf k} + {\bf q}|^3} \right] \,, \nonumber \\
&& \label{1lb}
\end{eqnarray}
where the $\ln(kL) \sim \mathcal{O}(1)$ factor comes from the evaluation of the momentum integrals in a minimal box \cite{lyth1,bernardeu4}, the $\mathcal{M}_3 (k_1,k_2,k_3)$ function is defined by \cite{seery5}
\begin{equation}
\mathcal{M}_3 (k_1,k_2,k_3) = -k_1 k_2^2 - 4 \frac{k_2^3 k_3^3}{k_t} + \frac{1}{2} k_1^3 + \frac{k_2^2 k_3^2}{k_t^2} (k_2-k_3) \,, 
\end{equation}
with $k_t = k_1+k_2+k_3$,
and the subindex $perm. \ l2a.$ means a permutation over the last two arguments in $\mathcal{M}_3$.

A quick glance reveals that the first term in Eq. (\ref{1la}) is subleading with respect to the second one because $|\eta_\sigma| > |\eta_\phi|$ and $\exp[4N(\eta_\phi - \eta_\sigma)] \gg 1$. The same is true for Eq. (\ref{1lb}) where $\exp[2N(\eta_\phi - \eta_\sigma)] \gg 1$. Now, by comparing the orders of magnitude of the leading terms in Eqs. (\ref{1la}) and (\ref{1lb}), we conclude that:
\begin{eqnarray}
\frac{P_\zeta^{1-loop \; a}}{P_\zeta^{1-loop \; b}} &\sim& \frac{\frac{\eta_\sigma^2}{\eta_\phi^4 \phi_\star^4} \exp \left[4N (\eta_\phi - \eta_\sigma)\right] \left(\frac{H_\star}{2\pi}\right)^4 \frac{2\pi^2}{k^3}}{\frac{\eta_\sigma}{\eta_\phi^3 \phi_\star^3} \exp \left[2N (\eta_\phi - \eta_\sigma)\right] \left(\frac{H_\star}{2\pi}\right)^4 \frac{\epsilon_\phi^{1/2}}{m_P} \frac{2\pi^2}{k^3}} \nonumber \\
&=& \frac{\eta_\sigma}{\eta_\phi} \frac{m_P}{\phi_\star} \exp \left[2N(\eta_\phi - \eta_\sigma)\right] \frac{1}{\epsilon_\phi^{1/2}} \gg 1 \,,
\end{eqnarray}
where $m_P \gg \phi_\star$ and $\epsilon_\phi \ll 1$. Thus, the one-loop leading term for $P_\zeta$ in our model is given by
\begin{equation}
P_\zeta^{1-loop} =\frac{2\pi^2}{k^3}  \frac{\eta_\sigma^2}{\eta_\phi^4 \phi_\star^4} \exp\left[4N(\eta_\phi - \eta_\sigma)\right]  \left(\frac{H_\star}{2\pi}\right)^4 \ln(kL) \,. \label{1lfpd}
\end{equation}

Having presented the leading tree-level and one-loop contributions to $P_\zeta$ in Eqs. (\ref{pt1}) and (\ref{1lfpd}), a consistency issue to think about is the dependence of the expression in Eq. (\ref{q1l}) on the infrared cutoff $L$. This quantity is in principle an artefact of the series expansion, and the final series result should in principle be independent on the chosen value for $L$ (see for instance Ref. \cite{riotto}).  In fact, by assuming that this is the case, Refs. \cite{lyth1,bartolo,enqvist2} have shown that there is a running on the $N$ derivatives with respect to $L$ so that changes in the $\ln (kL)$ factors are compensated by the running of the $N$ derivatives. This is similar to what happens in Quantum Field Theory where physical results independent on the energy scale must be independent of the chosen value for the renormalisation scale $Q$.  Changing $Q$ only modifies the relative weight of the tree-level and loop contributions, usually making the tree-level terms dominate over the loop corrections if $Q$ is chosen around the relevant energy scale of the process studied.  Nevertheless, we see that the $\ln (kL)$ term in Eq. (\ref{q1l}) does not compensate for the $\ln(kL)$ term in Eq. (\ref{1lfpd}), 
which is a real concern as we could expect since $\zeta$ and its spectral functions are a set of observables. The solution to this paradox relies on the fact that the observed $\zeta$ depends on $L$ as the stochastic properties of the distributions depend on the size of the available region in which we are actually able to perform observations. In this regard $\zeta$ is analogous to for instance the fine structure constant in Quantum Field Theory which, being an observable, depends on the energy scale for which experiments are done and, therefore, on $Q$. Likewise, $\zeta$ and its spectral functions, though being observables, depend on the size of the regions where observations are done and, therefore, on $L$. Having this in mind it is essential to work in a minimal box \cite{bartolo}, i.e. with $L$ a bit bigger than $H_0^{-1}$ (with the subscript 0 meaning today), so that $\ln (kL) \sim \mathcal{O} (1)$ as has been done throughout this paper. 

\subsection{Tree-level diagrams for $B_\zeta$}

\begin{figure}
\begin{center}
\begin{tabular}{cccc}
\includegraphics[width=7cm,height=2cm]{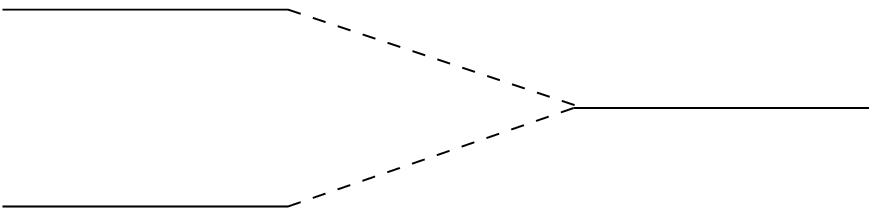} & & & \includegraphics[width=7cm,height=2cm]{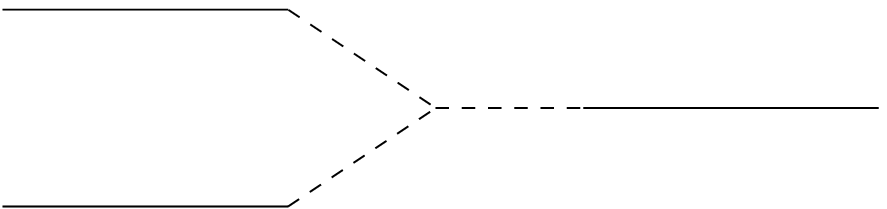} \\
(a) & & & (b)
\end{tabular}
\end{center}
\caption{Tree-level Feynman-like diagrams for $B_\zeta$. (a). The two internal dashed lines correspond to two-point correlators of field perturbations. (b). The internal dashed lines correspond to a three-point correlator of field perturbations.} \label{tlbd}
\end{figure}

Looking at Figs. \ref{tlbd}a and \ref{tlbd}b, we see that $B_\zeta^{tree}$ is given by two contributions $B_\zeta^{tree \; a}$ and $B_\zeta^{tree \; b}$:
\begin{eqnarray}
B_\zeta^{tree \; a} &=& N_\phi^2 N_{\phi \phi} \left[P_{\delta \phi} (k_1) \ P_{\delta \phi} (k_2) + 2 \ {\rm permutations} \right] \nonumber \\
&=& - \frac{1}{\eta_\phi^3 \phi_\star^4} \left(\frac{\sum_i k_i^3}{\prod_i k_i^3}\right) 4\pi^4 \left(\frac{H_\star}{2\pi}\right)^4 \,. \label{tlba} \\
B_\zeta^{tree \; b} &=& N_\phi^3 B_{\delta \phi \ \delta \phi \ \delta \phi} (k_1,k_2,k_3) \nonumber \\
&=& \frac{1}{\eta_\phi^3 \phi_\star^3} 4\pi^4 \sum_{perm} \left(\frac{H_\star}{2\pi}\right)^4 \frac{\epsilon_\phi^{1/2}}{2\sqrt{2}m_P} \frac{\mathcal{M}_3(k_1,k_2,k_3)}{\prod_i k_i^3} \,. \label{tlbb}
\end{eqnarray}

Now, from comparing the order of magnitude of the expressions in Eqs. (\ref{tlba}) and (\ref{tlbb}), we conclude that:
\begin{eqnarray}
\frac{B_\zeta^{tree \; a}}{B_\zeta^{tree \; b}} &\sim& \frac{\frac{1}{\eta_\phi^3 \phi_\star^4} \left(\frac{\sum_i k_i^3}{\prod_i k_i^3}\right) 4\pi^4 \left(\frac{H_\star}{2\pi}\right)^4}{\frac{1}{\eta_\phi^3 \phi_\star^3} 4\pi^4 \sum_{perm} \left(\frac{H_\star}{2\pi}\right)^4 \frac{\epsilon_\phi^{1/2}}{m_P} \left(\frac{\sum_i k_i^3}{\prod_i k_i^3}\right)} \nonumber \\
&=& \frac{m_P}{\phi_\star} \frac{1}{\epsilon_\phi^{1/2}} \gg 1 \,,
\end{eqnarray}
which in fact is usual as demonstrated in Refs. \cite{vernizzi,lyth7}. Thus, the tree-level leading term for $B_\zeta$ in our model is given by:
\begin{equation}
B_\zeta^{tree} = - \frac{1}{\eta_\phi^3 \phi_\star^4} \left(\frac{H_\star}{2\pi}\right)^4  4\pi^4 \left(\frac{\sum_i k_i^3}{\prod_i k_i^3}\right)\,. \label{tlbd1}
\end{equation}

As was done for $P_\zeta$ in Subsection \ref{diagramsP}, 
the one-loop correction to the spectrum of the field perturbations must be taken into account for the sake of consistency when calculating the contribution associated to the diagram in Fig. \ref{tlbd}a.  The discussion about the relevance of this quantum one-loop correction is actually the same as in Subsection \ref{diagramsP} and, therefore, we may conclude with some confidence that the expression in Eq. (\ref{tlba}) is reliable.  As regards the diagram in Fig. \ref{tlbd}b, it is necessary to include the one-loop correction the three-point correlator of the field perturbations in Eq. (\ref{tlbb}), which in fact nobody has calculated yet even for the single-field case.  Nevertheless we might conjecture that, analogously to that for the $P_\zeta$ case, such a correction is negligible compared to the tree-level contribution to $B_\zeta$ and, therefore, the expression in Eq. (\ref{tlbb}) will also be reliable.

\subsection{One-loop diagrams for $B_\zeta$}

\begin{figure}
\begin{center}
\begin{tabular}{cccc}
\includegraphics[width=7cm,height=2cm]{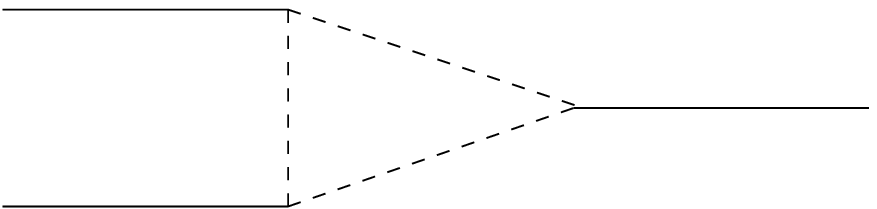} & & & \includegraphics[width=7cm,height=2cm]{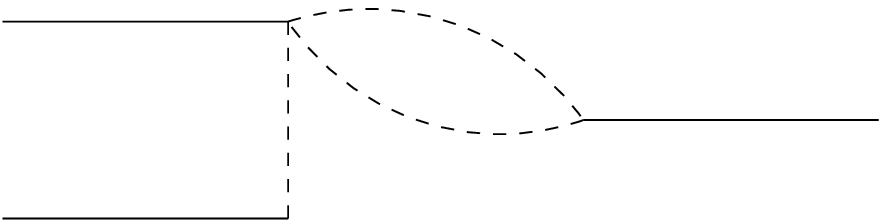} \\
(a) & & & (b) \\
& & & \\
& & & \\
\includegraphics[width=7cm,height=2cm]{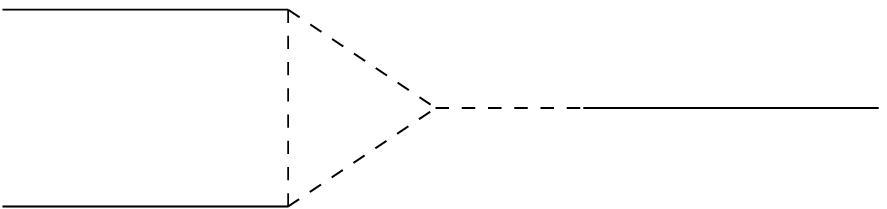} & & & \includegraphics[width=7cm,height=2cm]{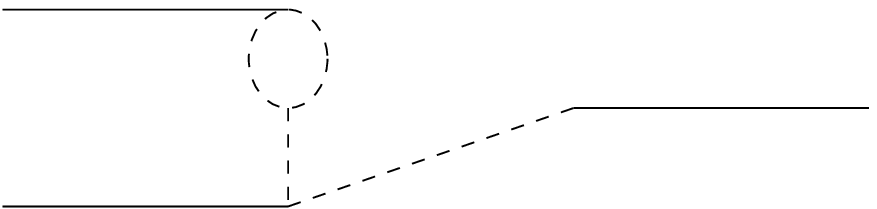} \\
(c) & & & (d) \\
& & & \\
& & & \\
\includegraphics[width=7cm,height=2cm]{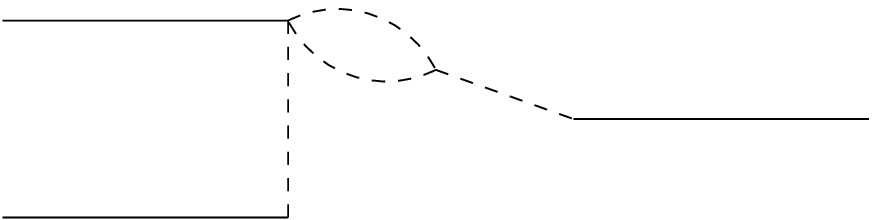} & & & \includegraphics[width=7cm,height=2cm]{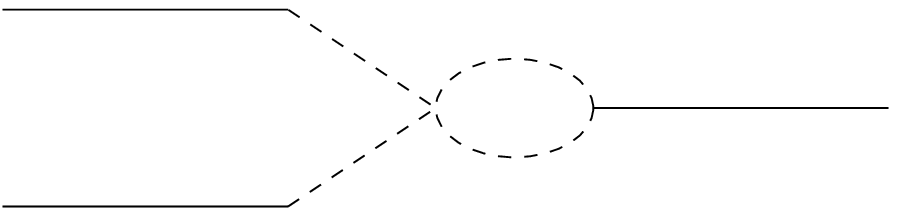} \\
(e) & & & (f)
\end{tabular}
\end{center}
\caption{One-loop Feynman-like diagrams for $B_\zeta$. (a) and (b). The three internal dashed lines correspond to two-point correlators of field perturbations. (c), (d), and (e). The internal dashed lines correspond to a two-point and a three-point correlator of field perturbations. (f). The internal dashed lines correspond to a four-point correlator of field perturbations.} \label{olbd}
\end{figure}

Looking at Figs. \ref{olbd}a, \ref{olbd}b, \ref{olbd}c, \ref{olbd}d, \ref{olbd}e, and \ref{olbd}f, we see that $B_\zeta^{tree}$ is given by six contributions $B_\zeta^{1-loop \; a}$, $B_\zeta^{1-loop \; b}$, $B_\zeta^{1-loop \; c}$, $B_\zeta^{1-loop \; d}$, $B_\zeta^{1-loop \; e}$, and $B_\zeta^{1-loop \; f}$:
\begin{eqnarray}
B_\zeta^{1-loop \; a} &=& \left[N_{\phi \phi}^3 + N_{\sigma \sigma}^3\right] \int \frac{d^3 q}{(2\pi)^3} P_{\delta \phi}(q) P_{\delta \phi} (|{\bf k_1} + {\bf q}|) P_{\delta \phi} (|{\bf k_3} - {\bf q}|) \nonumber \\
&=& \left[-\frac{1}{\eta_\phi^3 \phi_\star^6} + \frac{\eta_\sigma^3}{\eta_\phi^6 \phi_\star^6} \exp\left[6N (\eta_\phi - \eta_\sigma)\right] \right] \left(\frac{\sum_i k_i^3}{\prod_i k_i^3}\right) \ln(kL) \left(\frac{H_\star}{2\pi}\right)^6 4\pi^4 \,.  \label{poskomb} \\
B_\zeta^{1-loop \; b} &=& \frac{1}{2} \left[N_\phi N_{\phi \phi} N_{\phi \phi \phi} + N_\phi N_{\sigma \sigma} N_{\sigma \sigma \phi} \right] \times \nonumber \\
&&\times \left[\int \frac{d^3 q}{(2\pi)^3} P_{\delta \phi}(q) P_{\delta \phi}(|{\bf k_3} - {\bf q}|) P_{\delta \phi} (k_2) + 5 \ {\rm permutations}\right] \nonumber \\
&=& \frac{1}{2} \left[-\frac{2}{\eta_\phi^3 \phi_\star^6} - \frac{2\eta_\sigma^3}{\eta_\phi^6 \phi_\star^6} \exp \left[4N (\eta_\phi - \eta_\sigma)\right] \right] 16\pi^4 \left(\frac{\sum_i k_i^3}{\prod_i k_i^3}\right) \ln(kL) \left(\frac{H_\star}{2\pi}\right)^6 \,. \\
B_\zeta^{1-loop \; c} &=& N_\phi N_{\phi \phi}^2 \int \frac{d^3 q}{(2\pi)^3} \left[B_{\delta \phi \ \delta \phi \ \delta \phi} (q, |{\bf k_3} + {\bf q}|, k_3) P_{\delta \phi} (|{\bf k_1} - {\bf q}|) + 2 \ {\rm permutations} \right] + \nonumber \\
&& + N_\phi N_{\sigma \sigma}^2 \int \frac{d^3 q}{(2\pi)^3} \left[B_{\delta \sigma \ \delta \sigma \ \delta \phi} (q, |{\bf k_3} + {\bf q}|, k_3) P_{\delta \phi} (|{\bf k_1} - {\bf q}|) + 2 \ {\rm permutations} \right] \nonumber \\
&=& \frac{1}{\eta_\phi^3 \phi_\star^5} \Big[\int \frac{d^3 q}{(2\pi)^3} 8\pi^6 \sum_{perm} \left(\frac{H_\star}{2\pi}\right)^6 \frac{\epsilon_\phi^{1/2}}{2\sqrt{2}m_P} \frac{\mathcal{M}_3(q,|{\bf k_3} + {\bf q}|,k_3)}{q^3 |{\bf k_3} + {\bf q}|^3 k_3^3} \frac{1}{|{\bf k_1} - {\bf q}|^3} + \nonumber \\
&&+ 2 \ {\rm permutations} \Big] + \nonumber \\
&& + \frac{\eta_\sigma^2}{\eta_\phi^5 \phi_\star^5} \exp \left[4N (\eta_\phi - \eta_\sigma)\right] \times \nonumber \\
&& \times \Big[\int \frac{d^3 q}{(2\pi)^3} 8\pi^6 \sum_{perm. \ l2a.} \left(\frac{H_\star}{2\pi}\right)^6 \frac{\epsilon_\phi^{1/2}}{2\sqrt{2}m_P} \frac{\mathcal{M}_3(k_3,q,|{\bf k_3} + {\bf q}|)}{k_3^3 q^3 |{\bf k_3} + {\bf q}|^3} \frac{1}{|{\bf k_1} - {\bf q}|^3} + \nonumber \\
&& + 2 \ {\rm permutations} \Big] \,. \\
B_\zeta^{1-loop \; d} &=& \frac{1}{2} N_\phi N_{\phi \phi}^2 \int \frac{d^3 q}{(2\pi)^3} \left[B_{\delta \phi \ \delta \phi \ \delta \phi} (k_3, q, |{\bf k_3} - {\bf q}|) P_{\delta \phi} (k_2) + 5 \ {\rm permutations} \right] + \nonumber \\
&& + \frac{1}{2} N_\phi N_{\phi \phi} N_{\sigma \sigma} \int \frac{d^3 q}{(2\pi)^3} \left[B_{\delta \phi \ \delta \sigma \ \delta \sigma} (k_3, q, |{\bf k_3} - {\bf q}|) P_{\delta \phi} (k_2) + 5 \ {\rm permutations} \right] \nonumber \\
&=& \frac{1}{2\eta_\phi^3 \phi_\star^5} \Big[\int \frac{d^3 q}{(2\pi)^3} 8\pi^6 \sum_{perm} \left(\frac{H_\star}{2\pi}\right)^6 \frac{\epsilon_\phi^{1/2}}{2\sqrt{2}m_P} \frac{\mathcal{M}_3(k_3,q,|{\bf k_3} - {\bf q}|)}{k_3^3 q^3 |{\bf k_3} - {\bf q}|^3} \frac{1}{k_2^3} + \nonumber \\
&& + 5 \ {\rm permutations} \Big] - \nonumber \\
&& - \frac{\eta_\sigma}{2\eta_\phi^4 \phi_\star^5} \exp \left[2N (\eta_\phi - \eta_\sigma)\right] \times \nonumber \\
&& \times \Big[\int \frac{d^3 q}{(2\pi)^3} 8\pi^6 \sum_{perm. \ l2a.} \left(\frac{H_\star}{2\pi}\right)^6 \frac{\epsilon_\phi^{1/2}}{2\sqrt{2}m_P} \frac{\mathcal{M}_3(k_3,q,|{\bf k_3} - {\bf q}|)}{k_3^3 q^3 |{\bf k_3} - {\bf q}|^3} \frac{1}{k_2^3} + \nonumber \\
&& + 5 \ {\rm permutations} \Big] \,. \\
B_\zeta^{1-loop \; e} &=& \frac{1}{2} N_\phi^2 N_{\phi \phi \phi} \int \frac{d^3 q}{(2\pi)^3} \left[B_{\delta \phi \ \delta \phi \ \delta \phi} (k_1, q, |{\bf k_1} + {\bf q}|) P_{\delta \phi} (k_2) + 5 \ {\rm permutations} \right] + \nonumber \\
&& + \frac{1}{2} N_\phi^2 N_{\sigma \sigma \phi} \int \frac{d^3 q}{(2\pi)^3} \left[B_{\delta \phi \ \delta \sigma \ \delta \sigma} (k_1, q, |{\bf k_1} + {\bf q}|) P_{\delta \phi} (k_2) + 5 \ {\rm permutations} \right] \nonumber \\
&=& \frac{1}{\eta_\phi^3 \phi_\star^5} \Big[\int \frac{d^3 q}{(2\pi)^3} 8\pi^6 \sum_{perm} \left(\frac{H_\star}{2\pi}\right)^6 \frac{\epsilon_\phi^{1/2}}{2\sqrt{2}m_P} \frac{\mathcal{M}_3(k_1,q,|{\bf k_1} + {\bf q}|)}{k_1^3 q^3 |{\bf k_1} + {\bf q}|^3} \frac{1}{k_2^3} + \nonumber \\
&&+ 5 \ {\rm permutations} \Big] - \nonumber \\
&& - \frac{\eta_\sigma^2}{\eta_\phi^5 \phi_\star^5} \exp \left[2N (\eta_\phi - \eta_\sigma)\right] \times \nonumber \\
&& \times \Big[\int \frac{d^3 q}{(2\pi)^3} 8\pi^6 \sum_{perm. \ l2a.} \left(\frac{H_\star}{2\pi}\right)^6 \frac{\epsilon_\phi^{1/2}}{2\sqrt{2}m_P} \frac{\mathcal{M}_3(k_1,q,|{\bf k_1} + {\bf q}|)}{k_1^3 q^3 |{\bf k_1} + {\bf q}|^3} \frac{1}{k_2^3} + \nonumber \\
&& + 5 \ {\rm permutations} \Big] \,. \\
B_\zeta^{1-loop \; f} &=& \frac{1}{2} N_\phi^2 N_{\phi \phi} \int \frac{d^3 q}{(2\pi)^3} \left[T_{\delta \phi \ \delta \phi \ \delta \phi \ \delta \phi} ({\bf k_1}, {\bf q}, {\bf k_3} - {\bf q}, {\bf k_2}) + 2 \ {\rm permutations} \right] + \nonumber \\
&& + \frac{1}{2} N_\phi^2 N_{\sigma \sigma} \int \frac{d^3 q}{(2\pi)^3} \left[T_{\delta \phi \ \delta \sigma \ \delta \sigma \ \delta \phi} ({\bf k_1}, {\bf q}, {\bf k_3} - {\bf q}, {\bf k_2}) + 2 \ {\rm permutations} \right] \nonumber \\
&=& - \frac{1}{2\eta_\phi^3 \phi_\star^4} \Big[\int \frac{d^3 q}{(2\pi)^3} 8\pi^6 \sum_{perm} \left(\frac{H_\star}{2\pi}\right)^6 \frac{\mathcal{M}_4({\bf k_1},{\bf q},{\bf k_3} - {\bf q}, {\bf k_2})}{k_1^3 q^3 |{\bf k_3} - {\bf q}|^3 k_2^3} \frac{1}{m_P^2} + \nonumber \\
&&+ 2 \ {\rm permutations} \Big] + \nonumber \\
&& + \frac{\eta_\sigma}{2\eta_\phi^4 \phi_\star^4} \exp \left[2N (\eta_\phi - \eta_\sigma)\right] \times \nonumber \\
&& \times \Big[\int \frac{d^3 q}{(2\pi)^3} 8\pi^6 \sum_{perm. \ f2a. \ l2a.} \left(\frac{H_\star}{2\pi}\right)^6 \frac{\mathcal{M}_4({\bf k_1},{\bf k_2},{\bf q},{\bf k_3} - {\bf q})}{k_1^3 k_2^3 q^3 |{\bf k_3} - {\bf q}|^3} \frac{1}{m_P^2} + \nonumber \\
&& + 2 \ {\rm permutations} \Big] \,,
\end{eqnarray}
where the subindex $perm. \ f2a. \ l2a.$ means a permutation over the first two arguments and simultaneously over the last two arguments in $\mathcal{M}_4 ({\bf k_1},{\bf k_2},{\bf k_3},{\bf k_4})$ defined by \cite{seery4}
\begin{eqnarray}
\mathcal{M}_4 ({\bf k_1},{\bf k_2},{\bf k_3},{\bf k_4}) &=& -2 \frac{k_1^2 k_3^2}{k_{12}^2 k_{34}^2} \frac{W_{24}}{k_t} \left[\frac{{\bf Z}_{12}\cdot{\bf Z}_{34}}{k_{34}^2} + 2{\bf k}_2 \cdot {\bf Z}_{34} + \frac{3}{4} \sigma_{12} \sigma_{34} \right] \nonumber \\
&& -\frac{1}{2} \frac{k_3^2}{k_{34}^2} \sigma_{34} \left[\frac{{\bf k}_1 \cdot {\bf k}_2}{k_t} W_{124} + \frac{k_1^2 k_2^2}{k_t^3} \left(2 + 6\frac{k_4}{k_t}\right) \right] \,,
\end{eqnarray}
with ${\bf k}_{ij} = {\bf k}_i + {\bf k}_j$, $k_t = k_1 + k_2 + k_3 + k_4$, and
\begin{eqnarray}
&&\sigma_{ij} = {\bf k}_i \cdot {\bf k}_j + k_j^2 \,, \\
&&{\bf Z}_{ij} = \sigma_{ij} {\bf k}_i - \sigma_{ji}{\bf k}_j \,, \\
&&W_{ij} = 1 + \frac{k_i + k_j}{k_t} + \frac{2k_ik_j}{k_t^2} \,, \\
&&W_{lmn} = 1 + \frac{k_l + k_m + k_n}{k_t} + \frac{2(k_lk_m + k_lk_n + k_mk_n)}{k_t^2} + \frac{6k_lk_mk_n}{k_t^3} \,.
\end{eqnarray}

Following the same kind of analysis as we carried out for the one-loop diagrams of $P_\zeta$ and the tree-level terms for $B_\zeta$ we conclude the following:
\begin{eqnarray}
\frac{B_\zeta^{1-loop \; a}}{B_\zeta^{1-loop \; b}} &\sim& \frac{\frac{\eta_\sigma^3}{\eta_\phi^6 \phi_\star^6} \exp\left[6N (\eta_\phi - \eta_\sigma)\right] \left(\frac{\sum_i k_i^3}{\prod_i k_i^3}\right) \left(\frac{H_\star}{2\pi}\right)^6 4\pi^4}{\frac{\eta_\sigma^3}{\eta_\phi^6 \phi_\star^6} \exp \left[4N (\eta_\phi - \eta_\sigma)\right] 4\pi^4 \left(\frac{\sum_i k_i^3}{\prod_i k_i^3}\right) \left(\frac{H_\star}{2\pi}\right)^6} \nonumber \\
&=& \exp \left[2N(\eta_\phi - \eta_\sigma)\right] \gg 1 \,,
\end{eqnarray}
\begin{eqnarray}
\frac{B_\zeta^{1-loop \; a}}{B_\zeta^{1-loop \; c}} &\sim& \frac{\frac{\eta_\sigma^3}{\eta_\phi^6 \phi_\star^6} \exp\left[6N (\eta_\phi - \eta_\sigma)\right] \left(\frac{\sum_i k_i^3}{\prod_i k_i^3}\right) \left(\frac{H_\star}{2\pi}\right)^6 4\pi^4}{\frac{\eta_\sigma^2}{\eta_\phi^5 \phi_\star^5} \exp \left[4N (\eta_\phi - \eta_\sigma)\right] 4\pi^4 \left(\frac{H_\star}{2\pi}\right)^6 \frac{\epsilon_\phi^{1/2}}{m_P} \left(\frac{\sum_i k_i^3}{\prod_i k_i^3}\right)} \nonumber \\
&=& \frac{\eta_\sigma}{\eta_\phi} \frac{m_P}{\phi_\star} \exp \left[2N(\eta_\phi - \eta_\sigma)\right] \frac{1}{\epsilon_\phi^{1/2}} \gg 1 \,,
\end{eqnarray}
\begin{eqnarray}
\frac{B_\zeta^{1-loop \; a}}{B_\zeta^{1-loop \; d}} &\sim& \frac{\frac{\eta_\sigma^3}{\eta_\phi^6 \phi_\star^6} \exp\left[6N (\eta_\phi - \eta_\sigma)\right] \left(\frac{\sum_i k_i^3}{\prod_i k_i^3}\right) \left(\frac{H_\star}{2\pi}\right)^6 4\pi^4}{\frac{\eta_\sigma}{\eta_\phi^4 \phi_\star^5} \exp \left[2N (\eta_\phi - \eta_\sigma)\right] 4\pi^4 \left(\frac{H_\star}{2\pi}\right)^6 \frac{\epsilon_\phi^{1/2}}{m_P} \left(\frac{\sum_i k_i^3}{\prod_i k_i^3}\right)} \nonumber \\
&=& \left(\frac{\eta_\sigma}{\eta_\phi}\right)^2 \frac{m_P}{\phi_\star} \exp \left[4N(\eta_\phi - \eta_\sigma)\right] \frac{1}{\epsilon_\phi^{1/2}} \gg 1 \,,
\end{eqnarray}
\begin{eqnarray}
\frac{B_\zeta^{1-loop \; a}}{B_\zeta^{1-loop \; e}} &\sim& \frac{\frac{\eta_\sigma^3}{\eta_\phi^6 \phi_\star^6} \exp\left[6N (\eta_\phi - \eta_\sigma)\right] \left(\frac{\sum_i k_i^3}{\prod_i k_i^3}\right) \left(\frac{H_\star}{2\pi}\right)^6 4\pi^4}{\frac{\eta_\sigma^2}{\eta_\phi^5 \phi_\star^5} \exp \left[2N (\eta_\phi - \eta_\sigma)\right] 4\pi^4 \left(\frac{H_\star}{2\pi}\right)^6 \frac{\epsilon_\phi^{1/2}}{m_P} \left(\frac{\sum_i k_i^3}{\prod_i k_i^3}\right)} \nonumber \\
&=& \frac{\eta_\sigma}{\eta_\phi} \frac{m_P}{\phi_\star} \exp \left[4N(\eta_\phi - \eta_\sigma)\right] \frac{1}{\epsilon_\phi^{1/2}} \gg 1 \,,
\end{eqnarray}
\begin{eqnarray}
\frac{B_\zeta^{1-loop \; a}}{B_\zeta^{1-loop \; f}} &\sim& \frac{\frac{\eta_\sigma^3}{\eta_\phi^6 \phi_\star^6} \exp\left[6N (\eta_\phi - \eta_\sigma)\right] \left(\frac{\sum_i k_i^3}{\prod_i k_i^3}\right) \left(\frac{H_\star}{2\pi}\right)^6 4\pi^4}{\frac{\eta_\sigma}{\eta_\phi^4 \phi_\star^4} \exp \left[2N (\eta_\phi - \eta_\sigma)\right] 4\pi^4 \left(\frac{H_\star}{2\pi}\right)^6 \frac{1}{m_P^2} \left(\frac{\sum_i k_i^3}{\prod_i k_i^3}\right)} \nonumber \\
&=& \left(\frac{\eta_\sigma}{\eta_\phi}\right)^2 \left(\frac{m_P}{\phi_\star}\right)^2 \exp \left[4N(\eta_\phi - \eta_\sigma)\right] \gg 1 \,.
\end{eqnarray}
Thus, the one-loop leading term for $B_\zeta$ in our model is given by:
\begin{equation}
B_\zeta^{1-loop} = \frac{\eta_\sigma^3}{\eta_\phi^6 \phi_\star^6} \exp\left[6N (\eta_\phi - \eta_\sigma)\right] \left(\frac{H_\star}{2\pi}\right)^6 \ln(kL) 4\pi^4 \left(\frac{\sum_i k_i^3}{\prod_i k_i^3}\right) \,. \label{1lfbd}
\end{equation}

Once again, the $\ln (kL)$ dependence in Eq. (\ref{1lfbd}) does not look like that obtained from introducing Eq. (\ref{q1l}) into Eq. (\ref{tlbd1}).  However the situation here is the same as that discussed at the end of Subsection \ref{Ldis}, leading us to identical conclusions.

\renewcommand{\refname}{{\large References}}


\begin{thebibliography}{30}


\bibitem{cobe}
NASA's COBE mission homepage: http://lambda.gsfc.nasa.gov/product/cobe/.

\bibitem{smooth}
G. F. Smooth {\it et. al.}, {\it Structure in the COBE Differential Microwave Radiometer First-Year Maps}, Astrophys. J. {\bf 396}, L1 (1992).

\bibitem{wmap}
NASA's WMAP mission homepage: http://wmap.gsfc.nasa.gov/.

\bibitem{hinshaw}
G. Hinshaw {\it et. al.}, {\it Five-Year Wilkinson Microwave Anisotropy Probe (WMAP) Observations: Data Processing, Sky Maps, \& Basic Results}, {\tt arXiv:0803.0732 [astro-ph]}.

\bibitem{komatsu2}
E. Komatsu {\it et. al.}, {\it First-Year Wilkinson Microwave Anisotropy Probe (WMAP) Observations: Tests of Gaussianity}, Astrophys. J. Suppl. Ser. {\bf 148}, 119 (2003).

\bibitem{komatsu1}
E. Komatsu {\it et. al.}, {\it Five-Year Wilkinson Microwave Anisotropy Probe (WMAP) Observations: Cosmological Interpretation}, {\tt arXiv:0803.0547 [astro-ph]}.

\bibitem{planck}
ESA's PLANCK mission homepage: http://planck.esa.int/.

\bibitem{planck1}
The Planck Collaboration, {\it The Scientific Programme of Planck}, {\tt arXiv:astro-ph/0604069}.

\bibitem{komatsu}
E. Komatsu and D. N. Spergel, {\it Acoustic Signatures in the Primary Microwave Background Bispectrum}, Phys. Rev. D {\bf 63}, 063002 (2001).

\bibitem{starobinsky}
A. A. Starobinsky, {\it Multicomponent De Sitter (Inflationary) Stages and the Generation of Perturbations}, Pisma Zh. Eksp. Teor. Fiz. {\bf 42}, 124 (1985) [JETP Lett. {\bf 42}, 152 (1985)].

\bibitem{sasaki2}
M. Sasaki and E. D. Stewart, {\it A General Analytic Formula for the Spectral Index of the Density Perturbations Produced During Inflation}, Prog. Theor. Phys. {\bf 95}, 71 (1996).

\bibitem{lyth4}
D. H. Lyth, K. A. Malik, and M. Sasaki, {\it A General Proof of the Conservation of the Curvature Perturbation}, JCAP {\bf 0505}, 004 (2005).

\bibitem{lyth2}
D. H. Lyth and Y. Rodr\'{\i}guez, {\it Inflationary Prediction for Primordial Non-Gaussianity}, Phys. Rev. Lett. {\bf 95}, 121302 (2005).

\bibitem{boubekeur1}
L. Boubekeur and D. H. Lyth, {\it Detecting a Small Perturbation through its Non-Gaussianity}, Phys. Rev. D {\bf 73}, 021301(R) (2006).

\bibitem{alabidi2}
L. Alabidi and D. H. Lyth, {\it Inflation Models and Observation}, JCAP {\bf 0605}, 016 (2006).  See actually arXiv version: {\tt arXiv:astro-ph/0510441}.

\bibitem{zaballa}
I. Zaballa, Y. Rodr\'{\i}guez, and D. H. Lyth, {\it Higher Order Contributions to the Primordial Non-Gaussianity}, JCAP {\bf 0606}, 013 (2006).

\bibitem{alabidi1}
L. Alabidi, {\it Non-Gaussianity for a Two Component Hybrid Model of Inflation}, JCAP {\bf 0610}, 015 (2006).

\bibitem{vernizzi}
F. Vernizzi and D. Wands, {\it Non-Gaussianities in Two-Field Inflation}, JCAP {\bf 0605}, 019 (2006).

\bibitem{battefeld}
T. Battefeld and R. Easther, {\it Non-Gaussianities in Multi-Field Inflation}, JCAP {\bf 0703}, 020 (2007).

\bibitem{yokoyama1}
S. Yokoyama, T. Suyama, and T. Tanaka, {\it Primordial Non-Gaussianity in Multi-Scalar Slow-Roll Inflation}, JCAP {\bf 0707}, 013 (2007).

\bibitem{yokoyama2}
S. Yokoyama, T. Suyama, and T. Tanaka, {\it Primordial Non-Gaussianity in Multi-Scalar Inflation}, Phys. Rev. D {\bf 77}, 083511 (2008).

\bibitem{seery3}
D. Seery and J. E. Lidsey, {\it Non-Gaussianity from the Inflationary Trispectrum}, JCAP {\bf 0701}, 008 (2007).

\bibitem{byrnes2}
C. T. Byrnes, M. Sasaki, and D. Wands, {\it The Primordial Trispectrum from Inflation}, Phys. Rev. D {\bf 74}, 123519 (2006).

\bibitem{liddle}
A. R. Liddle and D. H. Lyth, {\it Cosmological Inflation and Large-Scale Structure}, Cambridge University Press, 2000.

\bibitem{lyth5}
D. H. Lyth and A. Riotto, {\it Particle Physics Models of Inflation and the Cosmological Density Perturbation}, Phys. Rep. {\bf 314}, 1 (1999).

\bibitem{lyth6}
D. H. Lyth, {\it Particle Physics Models of Inflation}, Lect. Notes Phys. {\bf 738}, 81 (2008).

\bibitem{maldacena}
J. Maldacena, {\it Non-Gaussian Features of Primordial Fluctuations in Single Field Inflationary Models}, JHEP {\bf 0305}, 013 (2003).

\bibitem{seery7}
D. Seery, K. A. Malik, and D. H. Lyth, {\it Non-Gaussianity of Inflationary Field Perturbations from the Field Equation}, JCAP {\bf 0803}, 014 (2008).

\bibitem{seery5}
D. Seery and J. E. Lidsey, {\it Primordial Non-Gaussianities from Multiple-Field Inflation}, JCAP {\bf 0509}, 011 (2005).

\bibitem{li}
S.-W. Li and W. Xue, {\it Revisiting Non-Gaussianity of Multiple-Field Inflation from the Field Equation}, {\tt arXiv:0804.0574 [astro-ph]}.

\bibitem{seery4}
D. Seery, J. E. Lidsey, and M. S. Sloth, {\it The Inflationary Trispectrum}, JCAP {\bf 0701}, 027 (2007).

\bibitem{bernardeu1}
F. Bernardeu and J.-P. Uzan, {\it Non-Gaussianity in Multi-Field Inflation}, Phys. Rev. D {\bf 66}, 103506 (2002).

\bibitem{bernardeu2}
F. Bernardeu and J.-P. Uzan, {\it Inflationary Models Inducing Non-Gaussian Metric Fluctuations}, Phys. Rev. D {\bf 67}, 121301 (2003).

\bibitem{lyth3}
D. H. Lyth and Y. Rodr\'{\i}guez, {\it Non-Gaussianity from the Second-Order Cosmological Perturbation}, Phys. Rev. D {\bf 71}, 123508 (2005).

\bibitem{cogollo}
H. R. S. Cogollo, Y. Rodr\'{\i}guez, and C. A. Valenzuela-Toledo, {\it On the Issue of the $\zeta$ Series Convergence and Loop Corrections in the Generation of Observable Primordial Non-Gaussianity in Slow-Roll Inflation. Part II: the Trispectrum}, to be submitted.

\bibitem{sasaki1}
M. Sasaki, J. V${\rm \ddot{a}}$liviita, and D. Wands, {\it Non-Gaussianity of the Primordial Perturbation in the Curvaton Model}, Phys. Rev. D {\bf 74}, 103003 (2006).

\bibitem{lyth8}
D. H. Lyth and D. Seery, {\it Classicality of the Primordial Perturbations}, Phys. Lett. B {\bf 662}, 309 (2008).

\bibitem{bernardeu3}
F. Bernardeu, T. Brunier, and J.-P. Uzan, {\it High Order Correlation Functions for Self Interacting Scalar Field in De Sitter Space}, Phys. Rev. D {\bf 69}, 063520 (2004).

\bibitem{sachs}
R. K. Sachs and A. M. Wolfe, {\it Perturbations of a Cosmological Model and Angular Variations of the Cosmic Microwave Background}, Astrophys. J. {\bf 147}, 73 (1967).

\bibitem{com}
E. Komatsu, {\it private communication}.

\bibitem{bunn}
E. F. Bunn and M. J. White, {\it The Four-Year COBE Normalization and Large-Scale Structure}, Astrophys. J. {\bf 480}, 6 (1997).

\bibitem{alabidi3}
L. Alabidi and D. H. Lyth, {\it Inflation Models after WMAP Year Three}, JCAP {\bf 0608}, 013 (2006).

\bibitem{okamoto}
T. Okamoto and W. Hu, {\it Angular Trispectra of CMB Temperature and Polarization}, Phys. Rev. D {\bf 66}, 063008 (2002).

\bibitem{kogo}
N. Kogo and E. Komatsu, {\it Angular Trispectrum of CMB Temperature Anisotropy from Primordial Non-Gaussianity with the Full Radiation Transfer Function}, Phys. Rev. D {\bf 73}, 083007 (2006).

\bibitem{cooray1}
A. Cooray, {\it 21-cm Background Anisotropies Can Discern Primordial Non-Gaussianity}, Phys. Rev. Lett. {\bf 97}, 261301 (2006).

\bibitem{cooray2}
A. Cooray, C. Li, and A. Melchiorri, {\it The Trispectrum of 21-cm Background Anisotropies as a Probe of Primordial Non-Gaussianity}, Phys. Rev. D {\bf 77}, 103506 (2008).

\bibitem{yadav}
A. P. S. Yadav and B. D. Wandelt, {\it Evidence of Primordial Non-Gaussianity ($f_{NL}$) in the Wilkinson Microwave Anisotropy Probe 3-Year Data at 2.8 $\sigma$}, Phys. Rev. Lett. {\bf 100}, 181301 (2008).

\bibitem{jeong}
E. Jeong and G. F. Smoot, {\it Probing Non-Gaussianity in the Cosmic Microwave Bacground Anisotropies: One Point Distribution Function}, {\tt arXiv:0710.2371 [astro-ph]}.

\bibitem{mukhanov}
V. F. Mukhanov, {\it Physical Foundations of Cosmology}, Cambridge University Press, 2005.

\bibitem{bardeen}
J. M. Bardeen, {\it Gauge Invariant Cosmological Perturbations}, Phys. Rev. D {\bf 22}, 1882 (1980).

\bibitem{bunch}
T. S. Bunch and P. C. W. Davies, {\it Quantum Field Theory in De Sitter Space: Renormalisation by Point Splitting}, Proc. R. Soc. Lond. A {\bf 360}, 117 (1978).

\bibitem{lyth1}
D. H. Lyth, {\it The Curvature Perturbation in a Box}, JCAP {\bf 0712}, 016 (2007).

\bibitem{bernardeu4}
F. Bernardeu and J.-P. Uzan, {\it Finite Volume Effects for Non-Gaussian Multi-Field Inflationary Models}, Phys. Rev. D {\bf 70}, 043533 (2004).

\bibitem{byrnes1}
C. T. Byrnes, K. Koyama, M. Sasaki, and D. Wands, {\it Diagrammatic Approach to Non-Gaussianity from Inflation}, JCAP {\bf 0711}, 027 (2007).

\bibitem{bartolo}
N. Bartolo, S. Matarrese, M. Pietroni, A. Riotto, and D. Seery, {\it On the Physical Significance of Infra-Red Corrections to Inflationary Observables}, JCAP {\bf 0801}, 015 (2008).

\bibitem{armendariz}
C. Armendariz-Picon, T. Damour, and V. Mukhanov, {\it k-inflation}, Phys. Lett. B {\bf 458}, 209 (1999).

\bibitem{silverstein}
E. Silverstein and D. Tong, {\it Scalar Speed Limits and Cosmology: Acceleration from D-cceleration}, Phys. Rev. D {\bf 70}, 103505 (2004).

\bibitem{arkani}
N. Arkani-Hamed, P. Creminelli, S. Mukohyama, and M. Zaldarriaga, {\it Ghost Inflation}, JCAP {\bf 0404}, 001 (2004).

\bibitem{seery6}
D. Seery and J. E. Lidsey, {\it Primordial Non-Gaussianities in Single Field Inflation}, JCAP {\bf 0506}, 003 (2005).

\bibitem{chen}
X. Chen, M.-X. Huang, S. Kachru, and G. Shiu, {\it Observational Signatures and Non-Gaussianities of General Single Field Inflation}, JCAP {\bf 0701}, 002 (2007).

\bibitem{gao}
X. Gao, {\it Primordial Non-Gaussianities of General Multiple-Field Inflation}, JCAP {\bf 0806}, 029 (2008).


\bibitem{langlois1}
D. Langlois, S. Renaux-Petel, D. A. Steer, and T. Tanaka, {\it Primordial Fluctuations and Non-Gaussianities in Multi-Field DBI Inflation}, {\tt arXiv:0804.3139 [hep-th]}.

\bibitem{langlois2}
D. Langlois, S. Renaux-Petel, D. A. Steer, and T. Tanaka, {\it Primordial Perturbations and Non-Gaussianities in DBI and General Multi-Field Inflation}, {\tt arXiv:0806.0336 [hep-th]}.

\bibitem{arroja2}
F. Arroja, S. Mizuno, and K. Koyama, {\it Non-Gaussianity from the Bispectrum in General Multiple Field Inflation}, JCAP {\bf 0808}, 015 (2008).

\bibitem{arroja}
F. Arroja and K. Koyama, {\it Non-Gaussianity from the Trispectrum in General Single Field Inflation}, Phys. Rev. D {\bf 77}, 083517 (2008).

\bibitem{lyth9}
D. H. Lyth, {\it Generating the Curvature Perturbation at the End of Inflation}, JCAP {\bf 0511}, 006 (2005).

\bibitem{bernardeu5}
F. Bernardeu, L. Kofman, and J.-P. Uzan, {\it Modulated Fluctuations from Hybrid Inflation}, Phys. Rev. D {\bf 70}, 083004 (2004).

\bibitem{matsuda1}
T. Matsuda, {\it Modulated Inflation}, Phys. Lett. B {\bf 665}, 338 (2008).

\bibitem{matsuda2}
T. Matsuda, {\it Generating the Curvature Perturbation with Instant Preheating}, JCAP {\bf 0703}, 003 (2007).

\bibitem{sasakin1} M. Sasaki, {\it Multi-brid Inflation and Non-Gaussianity}, Prog. Theor. Phys. {\bf 120}, 159 (2008).

\bibitem{sasakin2} M. Sasaki, {\it A Note on Nonlinear Curvature Perturbations in an Exactly Soluble Model of Multi-Component Slow-Roll Inflation}, Class. Quantum Grav. {\bf 24}, 2433 (2007).

\bibitem{dodelson2}
S. Dodelson, W. H. Kinney, and E. W. Kolb, {\it Cosmic Microwave Background Measurements Can Discriminate among Inflation Models}, Phys. Rev. D {\bf 56}, 3207 (1997).

\bibitem{linde}
A. D. Linde, {\it  A New Inflationary Universe Scenario: A Possible Solution to the Horizon, Flatness, Homogeneity, Isotropy and Primordial Monopole Problems}, Phys. Lett. B {\bf 108}, 389 (1982).

\bibitem{albrecht}
A. Albrecht and P. J. Steinhardt, {\it  Cosmology for Grand Unified Theories with Radiatively Induced Symmetry Breaking}, Phys. Rev. Lett. {\bf 48}, 1220 (1982).

\bibitem{dimopoulos}
K. Dimopoulos and G. Lazarides, {\it  Modular Inflation and the Orthogonal Axion as Curvaton}, Phys. Rev. D {\bf 73}, 023525 (2006).

\bibitem{freese}
K. Freese, J. Frieman, and A. Olinto, {\it  Natural Inflation with Pseudo-Nambu-Goldstone Bosons}, Phys. Rev. Lett. {\bf 65}, 3233 (1990).

\bibitem{boubekeur2}
L. Boubekeur and D. H. Lyth, {\it Hilltop Inflation}, JCAP {\bf 0507}, 010 (2005).

\bibitem{ahmad}
I. Ahmad, Y.-S. Piao, and C.-F. Quiao, {\it The Spectrum of Curvature Perturbation for Multi-Field Inflation with a Small-Field Potential}, JCAP {\bf 0802}, 002 (2008).

\bibitem{linde2}
A. D. Linde, {\it Hybrid Inflation}, Phys. Rev. D {\bf 49}, 748 (1994).

\bibitem{enqvist1}
K. Enqvist and A. V${\rm \ddot{a}}$ihk${\rm \ddot{o}}$nen, {\it Non-Gaussian Perturbations in Hybrid Inflation}, JCAP {\bf 0409}, 006 (2004).

\bibitem{vaihkonen}
A. V${\rm \ddot{a}}$ihk${\rm \ddot{o}}$nen, {\it Comment on Non-Gaussianity in Hybrid Inflation}, {\tt arXiv:astro-ph/0506304}.

\bibitem{dodelson}
S. Dodelson, {\it Modern Cosmology}, Academic Press, 2003.

\bibitem{armendariz2}
C. Armendariz-Picon, M. Fontanini, R. Penco, and M. Trodden, {\it Where does Cosmological Perturbation Theory Break Down?}, {\tt arXiv:0805.0114 [hep-th]}.

\bibitem{rigopoulos}
G. Rigopoulos, E. P. S. Shellard, and B. J. W. van Tent, {\it Quantitative Bispectra from Multifield Inflation}, Phys. Rev. D {\bf 76}, 083512 (2007).

\bibitem{friedman} B. C. Friedman, A. Cooray, and A. Melchiorri, {\it WMAP-Normalized Inflationary Model Predictions and the Search for Primordial Gravitational Waves with Direct Detection Experiments}, Phys. Rev. D {\bf 74}, 123509 (2006).

\bibitem{creminelli}
P. Creminelli and M. Zaldarriaga, {\it Single Field Consistency Relation for the 3-Point Function}, JCAP {\bf 0410}, 006 (2004).

\bibitem{spivak} M. Spivak, {\it Calculus}, Cambridge University Press, 1994.

\bibitem{jarnhus}
P. R. Jarnhus and M. S. Sloth, {\it De Sitter Limit of Inflation and Nonlinear Perturbation Theory}, JCAP {\bf 0802}, 013 (2008).

\bibitem{seery2}
D. Seery, {\it One-Loop Corrections to the Curvature Perturbation from Inflation}, JCAP {\bf 0802}, 006 (2008).

\bibitem{weinberg1}
S. Weinberg, {\it Quantum Contributions to Cosmological Correlations}, Phys. Rev. D {\bf 72}, 043514 (2005).

\bibitem{weinberg2}
S. Weinberg, {\it Quantum Contributions to Cosmological Correlations. II. Can these Corrections Become Large?}, Phys. Rev. D {\bf 74}, 023508 (2006).

\bibitem{sloth1}
M. S. Sloth, {\it On the One-Loop Corrections to Inflation and the CMB Anisotropies}, Nucl. Phys. B {\bf 748}, 149 (2006).

\bibitem{sloth2}
M. S. Sloth, {\it On the One-Loop Corrections to Inflation II: The Consistency Relation}, Nucl. Phys. B {\bf 775}, 78 (2007).

\bibitem{seery1}
D. Seery, {\it One-Loop Corrections to a Scalar Field During Inflation}, JCAP {\bf 0711}, 025 (2007).

\bibitem{riotto}
A. Riotto and M. S. Sloth, {\it On Resumming Inflationary Perturbations Beyond One-Loop}, JCAP {\bf 0804}, 030 (2008).

\bibitem{enqvist2}
K. Enqvist, S. Nurmi, D. Podolsky, and G. I. Rigopoulos, {\it On the Divergences of Inflationary Superhorizon Perturbations}, JCAP {\bf 0804}, 025 (2008).

\bibitem{lyth7}
D. H. Lyth and I. Zaballa, {\it A Bound Concerning Primordial Non-Gaussianity}, JCAP {\bf 0510}, 005 (2005).

\end{thebibliography}
\end{document}